\documentclass[
  a4paper,%
  11pt,%
  extramargin,
  style=screen,
]{tubsartcl}

\newcommand{\clrLight}{tuBlueDark20}
\newcommand{\clrMedium}{tuBlueDark60}
\newcommand{\clrDark}{tuBlueDark80}
\newcommand{\clrImage}{img/mining}

\usepackage[utf8x]{inputenc}
\usepackage[english]{babel}
\usepackage{anyfontsize}
\usepackage{microtype}
\usepackage{sectsty}
\sectionfont{\color{\clrMedium}}
\subsectionfont{\color{\clrMedium}}

\usepackage{array}
\usepackage{algorithm}
\usepackage[noend]{algpseudocode}
\usepackage{amsmath}
\usepackage{amstext}
\usepackage{amssymb}
\usepackage{booktabs}
\usepackage{balance}
\usepackage{breakurl}
\usepackage{color}
\usepackage{enumerate}
\usepackage{enumitem}
\usepackage{eurosym}
\usepackage{fancyvrb}
\usepackage[T1]{fontenc}
\usepackage{graphicx}
\usepackage[breaklinks,hidelinks]{hyperref}
\usepackage{listings}
\usepackage{lmodern}
\usepackage{mathtools}
\usepackage{microtype}
\usepackage{multicol}
\usepackage{multirow}
\usepackage[square,sort&compress,numbers]{natbib}
\usepackage[detect-all,binary-units=true]{siunitx}
\usepackage{subfigure}
\usepackage{tikz}
\usepackage[textsize=scriptsize]{todonotes}
\usepackage{wasysym}
\usepackage{xspace}

\renewcommand{\paragraph}[1]{{\vskip 8pt \noindent \emph{#1.}}}

\SetMathAlphabet{\mathcal}{normal}{OMS}{lmsy}{m}{n}
\SetMathAlphabet{\mathcal}{bold}{OMS}{lmsy}{m}{n}
\SetSymbolFont{largesymbols}{normal}{OMX}{lmex}{m}{n}
\SetSymbolFont{largesymbols}{bold}{OMX}{lmex}{m}{n}

\definecolor{hlcolor}{gray}{0.92}

\newcommand{\empirical}[1]{\setlength{\fboxsep}{1pt}\fbox{#1}}
\renewcommand{\empirical}[1]{#1}

\DeclareRobustCommand{\officialeuro}{%
  \ifmmode\expandafter\text\fi
  {\fontencoding{U}\fontfamily{eurosym}\selectfont e}}

\definecolor{tubsRed}{cmyk}{0.1,1.0,0.8,0.0}

\usetikzlibrary{shapes,arrows,chains,positioning,calc}

\tikzset{
    start chain = going right,
    line/.style = {-latex', draw},
    doubleline/.style = {latex'-latex', double, draw},
    block/.style = {rectangle, draw, rounded corners, fill=tubsRed!20,
            text width=6em, align=center, minimum height=3em,
            on chain}
}

\newcommand{\perc}[1]{\SI{#1}{\percent}}

\newcommand{\tablesize}{\footnotesize}

\newcommand{\eg}{e.g.,\xspace} %

\newcommand{\code}[1]{{\small\texttt{#1}}}

\newcommand{\skipcite}[1]{\unskip\unpenalty}
\newcommand{\skipcitep}[1]{\skipcite{}}
\newcommand{\skipcitet}[1]{\skipcite{}}

\let\oldcitet\citet
\renewcommand{\citet}[1]{\mbox{\oldcitet{#1}}}

\newcommand{\tab}[1]{\hyperref[#1]{Table~\ref{#1}}}
\newcommand{\fig}[1]{\hyperref[#1]{Figure~\ref{#1}}}
\newcommand{\figs}[1]{\hyperref[#1]{Figures~\ref{#1}}}
\newcommand{\sect}[1]{\hyperref[#1]{Section~\ref{#1}}}
\newcommand{\sects}[1]{\hyperref[#1]{Sections~\ref{#1}}}
\newcommand{\appx}[1]{\hyperref[#1]{Appendix~\ref{#1}}}

\makeatletter
\newcommand{\footnoteref}[1]{\protected@xdef\@thefnmark{\ref{#1}}\@footnotemark}
\makeatother

\newcommand{\ourshorttitle}{Web-based Cryptojacking in the Wild}
\newcommand{\ourtitle}{\ourshorttitle}

\newcommand{\chris}{Christian Wressnegger}
\newcommand{\knrd}{Konrad Rieck}
\newcommand{\marius}{Marius Musch}
\newcommand{\martin}{Martin Johns}

\newcommand{\appsec}{Institute for Application Security\xspace}

\newcommand{\Cryptojacking}{Cryptojacking\xspace}
\newcommand{\cryptojacking}{cryptojacking\xspace}

\newcommand{\cryptojacker}{cryptojacker\xspace}
\newcommand{\cryptojackers}{cryptojackers\xspace}
\newcommand{\coinhive}{CoinHive\xspace}
\newcommand{\similarweb}{SimilarWeb\xspace}
\newcommand{\hashrate}{\SI{80}{H/s}\xspace}
\newcommand{\chpayout}{0.00005749\,XMR~per \num{1}~million hashes\xspace}
\newcommand{\phrevenue}{\SI{223.1}{XMR}\xspace}

\newcommand{\authors}{\marius, \chris, \\
\martin, and \knrd}

\title{\vspace{-25mm}}

\begin{document}

\begin{titlepage}
\showtubslogo
\showlogo{\includegraphics[width=5.6cm]{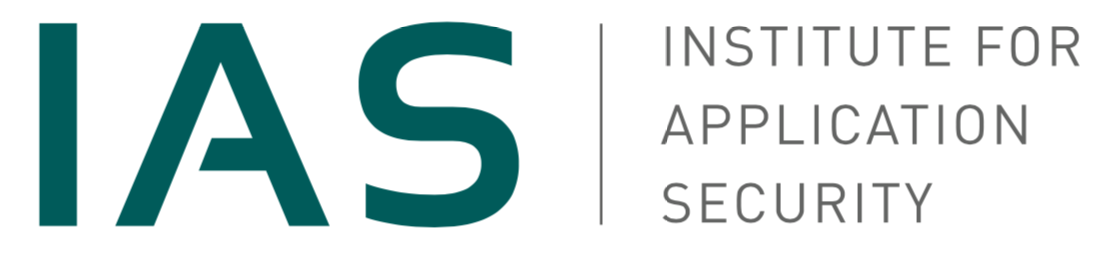}}

\begin{titlerow}[bgimage=\clrImage]{2}
\end{titlerow}

\begin{titlerow}[bgcolor=\clrDark, fgcolor=tubsWhite]{1}
  \raggedright \noindent \fontsize{24}{28}\selectfont
  \ourtitle
\end{titlerow}

\begin{titlerow}[bgcolor=\clrLight]{4}
  \raggedright \noindent \fontsize{16}{20}\selectfont
  \authors\\
  \vspace{6mm}
  Computer Science Report \\
  Technische Universität Braunschweig\\
  \appsec
\end{titlerow}
\end{titlepage}

\newpage
~
\vfill \noindent
Technische Universität Braunschweig\\
\appsec\\
Mühlenpfordtstraße 23\\
38106 Braunschweig, Germany\\
\newpage

\date{}
\author{}

\maketitle

\begin{abstract}
\subsection*{Abstract}
With the introduction of memory-bound cryptocurrencies, such as
Monero, the implementation of mining code in browser-based JavaScript
has become a worthwhile alternative to dedicated mining rigs.
Based on this technology, a new form of
parasitic computing, widely called \emph{\cryptojacking} or drive-by
mining, has gained momentum
in the web. A \cryptojacking site abuses the computing resources of
its visitors to covertly mine for
cryptocurrencies. In this paper, we systematically explore this
phenomenon. For this, we propose a 3-phase analysis approach, which
enables us to identify mining scripts and conduct a
large-scale study on the prevalence of \cryptojacking in the
Alexa~1~million websites.  We find that \cryptojacking is common, with
currently 1~out of 500~sites hosting a mining script. Moreover,
we perform several secondary analyses to gain insight into the
\cryptojacking landscape, including
a measurement of code characteristics, an estimate of
expected mining revenue, and an evaluation of current blacklist-based
countermeasures.
\end{abstract}

\maketitle

\section{Introduction}

Cryptocurrencies, such as Bitcoin and Ether, have gained popularity in
the last years, as they provide an alternative to centrally controlled
fiat money and a profitable playground for financial speculation. A
basic building block of these currencies is the process of
\emph{mining}, in which a group of users solves computational puzzles
to validate transactions and generate new coins of the currency
\citep[see][]{Nak09}. Although the stability and long-term
perspectives of cryptocurrencies are not fully understood, they have
attracted large user communities that mine and trade coins in
different markets with considerable volume.  For example, Bitcoin
reached an all-time high of \num{19300} USD per coin in
December~2017~\cite{website:coinmarket}, resulting in a market value
comparable to major companies.

The mining of cryptocurrencies has been largely dominated by dedicated
hardware systems, such as GPU and ASIC mining rigs. This situation,
however, has started to change with the introduction of
\emph{memory-bound} cryptocurrencies, like Monero, Bytecoin, and
Electroneum. These currencies build on computational puzzles that are
memory intensive and thereby reduce the advantage of specific hardware
over commodity processors \citep[see][]{CryptoNight13, CryptoNote13}.
Consequently, the resulting currencies can be profitably mined on
regular computer systems and thus open the door for the widespread
application of cryptocurrency mining.

Unfortunately, this development has also attracted miscreants who have
discovered cryptocurrencies as a new means for generating profit. By
tricking users into unnoticeably running a miner on their computers,
they can utilize the available resources for generating revenue---a
strategy denoted as \emph{\cryptojacking} or drive-by
mining~\citep{MineSweeper}. A novel realization of this
strategy is injecting mining code into a website, such that the
browser of the victim mines during the website's visit. First
variants of these attacks have emerged with the availability of the
\coinhive miner in September~2017~\citep[][]{website:adguard,
  website:krebs}. This software cleverly combines recent web
technologies to implement a miner that efficiently operates on all
major browsers. Although originally developed for benign purposes,
\coinhive has been maliciously injected into several websites over the
last months~\citep[e.g.,][]{website:goodin18, website:degroot}. Recently,
a vulnerability in MikroTik routers has been used to inject a \cryptojacking
script into traffic passing through more than \num{200000} of these
routers~\citep{website:mikrotik}.

In this paper, we present a large-scale study on web-based
\cryptojacking. While previous work has anecdotally described this
phenomenon~\citep{EskLeoMurCla18} and discussed detection
approaches~\citep{MineSweeper},
we systematically investigate the prevalence of mining
scripts in the Alexa Top 1~million websites. To this end, we have
instrumented a browser to monitor the execution of code during the
visit of a website and spot indications of mining activity, such as an
unusual CPU utilization, the excessive repetition of functions and the
presence of suspicious scripts. %
Moreover,
we have traced back mining activity to individual wallets and API
keys, which allows us to make estimates about the
revenue of particular \cryptojackers. In summary, our study provides the
following key insights:\\

\textbf{(a) Web-based cryptojacking is not rare.}  We observe that
1~out of \num{500}~websites in the Alexa ranking contains a web-based miner
that immediately starts once the website is visited. These miners
target different cryptocurrencies, including Monero, Bytecoin, and
Electroneum.  While the JavaScript code driving the mining is diverse
and often obfuscated, we observe that almost all miners employ
similar WebAssembly code from the \coinhive project. We credit this
finding to the CryptoNote protocol~\cite{CryptoNote13} that is
implemented by the \coinhive miner and can support different
currencies with minor modifications.\\[-8pt]

\textbf{(b) Mining profits are moderate.}  Our analysis further
provides a glimpse at the ecosystem of current \cryptojacking. Several
attackers operate on different websites using the same wallet or API
key. In some cases, a single website mines multiple currencies in
parallel, likely due to a coincident infection with malicious code.
Based on the configuration of typical desktop computers and statistics
about website visits, we estimate the revenue generated by individual
miners in the Alexa ranking at a range of a few cents up to
\empirical{\SI{340}{USD}} per day under the current price of the
respective cryptocurrencies.\\[-8pt]

\textbf{(c) Existing defenses are insufficient.} We investigate the
effectivity of current defenses against \cryptojacking, such as
blacklists and browser extensions.  While these defenses provide
sufficient protection from known mining sites, such as \coinhive and
CryptoLoot, the underlying static detection patterns are ineffective
against customized variants of the mining code. We thus argue that
better protection from web-based mining is needed and, in addition to
static matching, also run-time analysis needs to be considered to
reliably track down mining activity. MineSweeper~\citep{MineSweeper}, for
instance, is a promising alternative that provides detection based on
characteristics of cryptomining code.\\[-8pt]

The remainder of the paper is organized as follows: We first review
the background of memory-bound cryptocurrencies and web-based mining
in \sect{sec:background}. We then present our empirical study on
\cryptojacking, where the prevalence of web-based miners is discussed
in \sect{sec:identification} and their characteristics are analyzed in
\sect{sec:analysis}.
In \sect{sec:case-studies} we provide case studies on the two largest
families of miners and discuss limitations of the study in
\sect{sec:limitations}.
Finally, related work is
presented in \ref{sec:related-work} and
\sect{sec:conclusion} concludes the
paper.

\section{Web-based Mining}
\label{sec:background}

Cryptocurrencies are a specific type of electronic money that provide
decentralized control using the concept of blockchains~\citep{Nak09}.
In contrast to other electronic currencies, individuals generate revenue
by solving computational puzzles and thereby validating transactions---a
process referred to as \emph{mining}. While mining of classic
cryptocurrencies requires specific hardware to be profitable, memory-bound
currencies and novel web standards have paved the way for effective mining
in web browsers. In the following, we review these changes and discuss
their impact on \cryptojacking.

\subsection{Memory-bound Cryptocurrencies}

Classic cryptocurrencies, such as Bitcoin and Ether, build on
proof-of-work functions (computational puzzles) that are
\emph{CPU-bound}, that is, the effectivity of mining mainly depends on
the available computing power~\cite{BonMilClaNar+15}. Hardware devices
designed for demanding computations, such as GPUs and ASICs, thus
provide a better mining performance than common CPUs. As a
consequence, profitable mining of classic cryptocurrencies has become
largely infeasible with regular desktop and mobile computer systems.

\paragraph{Memory-bound functions}
This situation has not been anticipated in the original design of the
first cryptocurrencies and violates the ``one-CPU-one-vote'' principle
underlying Bitcoin mining \citep{Nak09}. As a remedy, alternative
cryptocurrencies have been developed in the community that make use of
memory-bound functions for constructing computational puzzles. One
prominent example is the cryptographic mixing protocol
\emph{CryptoNote} \cite{CryptoNote13} and the corresponding
proof-of-work function \emph{CryptoNight} \cite{CryptoNight13}.
CrypoNight is a hash function that determines the hash value for an
input object by extensively reading and writing elements from a
\SI{2}{Megabyte} memory region. This intensive memory access bounds
the run-time of the function and moves the overall mining performance
from the computing resources to the available memory access
performance. As memory access is comparably fast on common CPUs due to
multi-level caching, CryptoNight and other memory-bound proof-of-work
functions provide the basis for alternative cryptocurrencies that can
be efficiently mined on regular desktop systems and hence are a
prerequisite for realizing web-based miners.

\paragraph{CryptoNote-based currencies}
The idea of memory-bound proof-of-work functions along with other
improvements over the original Bitcoin protocol has spawned a series
of novel cryptocurrencies, each forking the concept of
CryptoNote. Prominent examples are
Monero~\citep[XMR,][]{website:monero},
Bytecoin~\citep[BCN,][]{website:bytecoin}, and
Electroneum~\citep[ETN,][]{website:electroneum}, which reach a market
capitalization between \num{226}~million and \num{3.8}~billion
USD~\citep{website:coinmarket}. These currencies share the underlying
CryptoNote protocol and thus can be easily implemented with same code
base. Moreover, due to the concept of anonymous transactions they
provide more privacy than Bitcoin and may conceal the identity of senders and
receivers~\citep[see][]{MoeSosHei+18, KumCleFisTopSax17}. \\[-8pt]

Both properties---profitable mining on desktop systems and the
availability of different currencies following the same cryptographic
protocol---render these currencies an ideal target for web-based
mining. Furthermore, the increased privacy of transactions provides a
basis for conducting \cryptojacking over manipulated web
sites. According to our findings, CryptoNote-based currencies are currently
prevalent in web-based mining and play a major role in \cryptojacking as
detailed in  Section~\ref{sec:analysis}.

\subsection{Novel Web Standards}
\label{sec:novelwebstandards}

The decentralized nature of cryptocurrencies imposes constraints on
the capabilities of mining clients. First, the clients need to
efficiently communicate with each other to synchronize the solving of
puzzles. Second, the clients require programming primitives that
enable an optimal utilization of available hardware resources.

At a first glance, these requirements seem to contradict with classic
web technology, as the underlying HTTP protocol induces a non-trivial
overhead and supported scripting languages, such as JavaScript and
ActionScript, do not provide efficient primitives for low-level
programming.  However, browser vendors and the W3C have continuously
advanced web standards and developed additional functionalities.
In combination, \emph{WebSockets}, \emph{WebWorkers} and
\emph{WebAssembly} provide a fruitful ground for web-based mining of
cryptocurrencies.

\paragraph{WebSockets} The WebSocket protocol has been standardized as
additional browser functionality in 2011 \citep{rfc:6455} and is
supported by all major browsers as of now. The protocol enables
full-duplex communication from the browser to a web server with less
overhead than HTTP. From the network perspective, the protocol is a
classic application-layer protocol that operates on top of the
transport layer. From the web application's point of view, however,
WebSockets rather provide a transport protocol that enables transferring
arbitrary payloads. %

In the context of web-based mining, WebSockets allow the efficient
communication between miners through a web server and thus are an
integral part of currently available implementations (see
Section~\ref{sec:available-miners}). However, WebSockets are also used
in several other types of web applications, like chats and multiplayer games,
and thus represent only a weak indicator of mining activity.

\paragraph{WebWorkers} A second addition are so-called WebWorkers which
have been introduced in 2015~\citep{w3c:workers} and are also
supported by all major browsers. This programming primitive enables
JavaScript code to schedule multiple threads and conduct concurrent
computations in the background. While the original programming model
underlying JavaScript already supports event-driven concurrency,
orchestrating the available computing resources, such as multiple
cores, has been technically involved. This problem is alleviated with
WebWorkers, where the number of concurrent threads can be scaled with
the available processor cores easily.

Although WebWorkers are not strictly necessary for implementing
web-based mining, they allow for better utilizing the available
resources and thus can also be found in most implementations. For our
study, we hence consider the presence multiple of WebWorker threads as an
indicator for potential mining activity.

\paragraph{WebAssembly} The previous two functionalities ease the
communication and scheduling of web-based miners. Yet they are
not sufficient for realizing an efficient implementation, as the
underlying JavaScript code requires a costly interpretation within the
browser. This problem is addressed by the WebAssembly standard from
2017~\citep{w3c:wasm}. The standard proposes a low-level bytecode
language that is a portable target for compilation of high-level
languages, such as C/C++ and Rust. WebAssembly code, or Wasm code for
short, is executed on a stack-based virtual machine in the browser and
improves the execution as well as loading time over JavaScript
code~\citep{HaaRosSchTit+17}. WebAssembly is currently supported by
Chrome, Safari, Firefox and Edge\footnote{Statistics from
  \url{https://caniuse.com/\#feat=wasm}, May~2018}.

WebAssembly is a perfect match for implementing mining software, as it
enables compiling cryptographic primitives, such as specific hash
functions, from a high-level programming language to low-level code for
a browser. As an example, \fig{fig:wasm1} shows a simplified snippet
of C code from the cryptographic hash Skein \citep{Skein08}. The
corresponding WebAssembly code is presented in \fig{fig:wasm2} as raw
bytes and instructions. Note that the instructions do not contain any
registers, due to the stack-based design of the virtual machine. The
characteristic constant of the Skein hash, which here is encoded in
LEB128 format---a variable-length representation of integer
numbers~\citep[see][]{DWARF93}---is visible in line 5.

\begin{figure}[htbp]
\centering

\subfigure[Simplified C code snippet from the Skein hash.]{%
\begin{minipage}[b]{0.85\linewidth}
\VerbatimInput[fontsize=\small, frame=single, numbers=left,
    numbersep=4pt]%
    {listings/example-c.txt}
    \vspace{0mm}
\end{minipage}
\label{fig:wasm1}}

\subfigure[Corresponding WebAssembly code as raw
bytes and instructions.]{%
\begin{minipage}[b]{0.85\linewidth}
\VerbatimInput[fontsize=\small, frame=single, numbers=left,
    numbersep=4pt]%
    {listings/example-wasm.txt}
    \vspace{-0mm}
\end{minipage}
\label{fig:wasm2}}\vspace{-1mm}

\caption{Example of C and WebAssembly code.}
\label{fig:wasm}
\end{figure}

\subsection{Anatomy of a Web-based Miner}
\label{sec:available-miners}

\coinhive is the first implementation of mining software based on the
aforementioned developments. The software has been officially released in
September~2017 and has originally been developed for a popular image
board as an alternative payment mechanism~\citep{website:krebs}.
Several variants that, similar to \coinhive, all implement the
CryptoNote protocol have been developed ever since, including JSECoin
and CryptoLoot. Although these implementations differ in some
details, they share how WebSockets, WebWorkers and WebAssembly are
used for efficient web-based mining. In the following, we describe the
anatomy of such miners and how these technologies interconnect.

The miner itself is distributed via a single JavaScript file, which the
website's owner includes on the page along with a small snippet to
configure and start the mining process. The snippet and its
configuration may be further customized, \eg to not execute on mobile
devices, but at least requires a unique id that maps miners to
identities---in the case of \coinhive, so-called \emph{site-keys}---in
order to account payouts for calculated hashes. Due to this
additional indirection it usually is not possible to link miners,
identified by site-keys, to specific wallet addresses. Moreover, each
account may be associated with multiple site-keys, such that multiple
mining sites may in fact mine for the same wallet without disclosing the
fact to the public.

On startup, the miner instantiates the desired number of WebWorkers and
creates a WebSocket connection to the mining pool, where it registers
itself with the supplied site-key. In return the miner receives a job
represented by a blob and a target: The blob is similar to CryptoNote's
block identifier and contains the current block header, the hash of the
Merkle tree root that allows to securely link the mined block to the
previous block on the chain, as well as the number of transactions
included in the block~\citep{CryptoNote003}. The target, on the other
hand, is a value chosen by the pool and determines whether a found hash
is reported. Since the pool rewards calculated hashes rather than
mined blocks, it needs to reliably track the miner's contribution.
A target value of \code{ffffff00}, for instance, means that every hash
with two trailing zeros should be sent to the pool (hashes are
represented in little-endian format), alongside with a nonce used to
create that hash. In this example, the probability of finding a nonce
which results in such a hash is~$\frac{1}{256}$, such that the pool
credits \num{256}~hashes to the owner of the site-key. This way, only a
fraction of the calculated hashes needs to be transferred and validated.
Simultaneously, this prevents the pool's participants from cheating the
system. A~simplified example of this communication is shown in
\fig{fig:websocket-background}.

\begin{figure}[htbp]
\centering

\begin{minipage}[b]{0.85\linewidth}
\VerbatimInput[fontsize=\small, frame=single, numbers=left,
    numbersep=4pt]%
    {listings/websocket.txt}
    \vspace{0mm}
\end{minipage}

\caption{WebSocket communcation, simplified for brevity. An arrow to the right indicates messages sent from the browser to the mining pool.}
\label{fig:websocket-background}
\end{figure}

The calculation of the CryptoNight hashes at the core of the whole
mining process is implemented in WebAssembly to increase performance. The
corresponding code is usually included in the JavaScript code of the miner as a
binary blob and instantiated by each worker. If throttling is configured to use
less than \perc{100}~of the CPU for mining, the workers constantly monitor the
time consumed for each calculated hash and adjust the delay between hash
calculations. This, however, only allows for a rough approximation of the
desired load on the system.

\subsection{\Cryptojacking}
\label{sec:cryptojacking}

Web-based mining certainly has legit use-cases and may pose an
alternative to online advertisements as scheme of monetization.
Moreover, mining might even replace CAPTCHAs used for rate limitation by
requiring a proof-of-work. The anonymity offered by cryptocurrencies in
combination with simple deployment on the web, unfortunately, also
attracts actors with less noble goals. As the effort and cost of
including a miner in an existing website is negligible, all it needs is
access to a frequently visited website. Recently, a variety of incidents
involving mining scripts have been reported for popular
websites~\citep{website:browsealoud,website:wired}.

We define \cryptojacking as the practice of automatically starting a
web-based miner upon visiting a web page. For this, we neither consider
the disclosure of the mining process to the user nor the presence of an
opt-out mechanism relevant. We view a consent after the fact as an inadmissible
mode of operation, similarly to how the GDPR now requires a ``clear affirmative
action'' for tracking cookies in the EU~\citep{GDPR}. Miners that only run after
explicit consent by the user, such as as Authedmine and JSEcoin, are not
considered part of the problem and are thus not examined in our study. To
conclude that a website employs \cryptojacking, we further do not differentiate
between scripts added by the website's owner and scripts injected by a third
party by means of hacking the server or hijacking included scripts.

\section{Identification of Web-based Miners}
\label{sec:identification}

Based on the discussed background, we proceed to present our systematic
study of \cryptojacking on the web. The goal of this study is to
evaluate to which degree the recent level of hype is justified through
painting a comprehensive picture of the current \cryptojacking practices
in the wild. To this end, we measure the prevalence of \cryptojacking in
today's web~(\sect{sec:prevalence}) and examine the effectiveness
of the current generation of dedicated anti-\cryptojacking
countermeasures~(\sect{sec:blacklists}).
Finally, we explore what evidence can be drawn from sites that host
\cryptojacking scripts (\sect{sec:distribution}).
After introducing our approach and documenting our experiments, we
address these topics in detail.

\subsection{General Approach}
\label{sec:general-approach}

\begin{figure*}[htbp]
  \centering
  \includegraphics[width=1.0\textwidth]{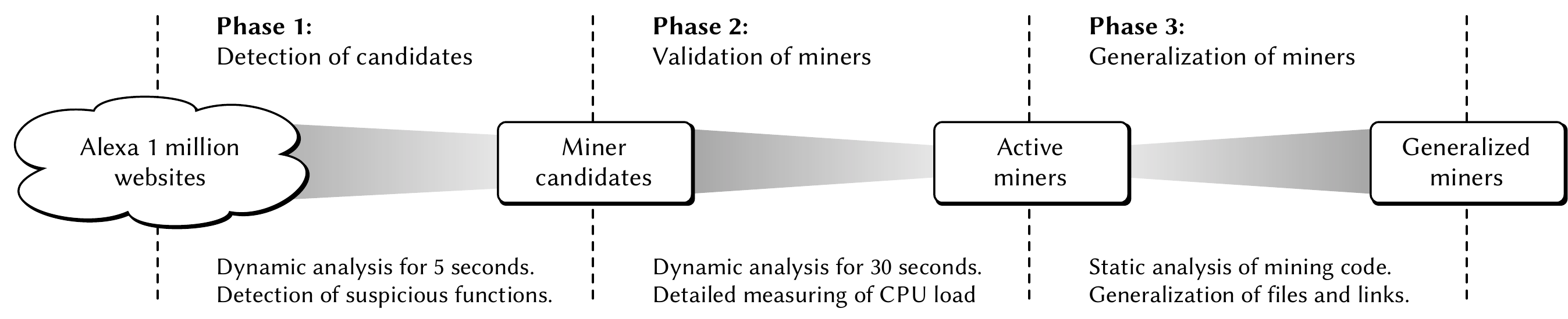}
  \caption{Overview of our approach for identification of web-based miners.}
  \label{fig:3phase}
\end{figure*}

For conducting our study, an empirical method is required that is
accurate in its detection capabilities while being scalable to enable
the analysis of large numbers of real-world websites. To accommodate
these requirements, we designed a dedicated \cryptojacking detection
process that spans three individual phases: 1)~An over-permissive first
broad sweep to identify potential miner \textit{candidates} using
heuristics, 2)~a thorough run-time analysis to isolate the real miners
within the candidate set, and 3)~a generalization step, in which we
extract static indicators, that allow the identification of
non-active or stealth mining scripts (see
\fig{fig:3phase}).

\paragraph{Phase 1: Detection of candidate sites}

In the first phase, our approach conducts a fast and imprecise initial
analysis of websites to create a pool of candidates which likely---but
not necessarily---host a mining script.
To do so, we compiled a set of heuristics that hint the potential
presence of a \cryptojacking script and that can be measured at
run-time while rendering a webpage in a browser. These heuristics were
extracted from a manual analysis of verified mining scripts.  For one, we
initiate a short profiling of the site's CPU usage, with unusual high CPU
utilization levels being interpreted as an indicator for mining. Furthermore,
we mark all sites as suspicious, that use miner-typical web technologies, which
are not in wide-spread use in the general web, namely WebAssembly or
non-trivial amounts of WebWorkers. If at least one of these indicators could be
found in a site, this site is marked as a potential \textit{mining candidate}.
Thus, the result of this phase is an over-approximation of the set of
actual mining sites.

\paragraph{Phase 2: Validation of mining scripts}

For obvious reasons, none of the used heuristics is conclusive in the
identification of miners, as there is a multitude of legitimate
reasons to use WebWorkers, Web\-Assembly or temporary high amounts of
computation. However, the constant and potentially unlimited usage of
CPU, caused by a \textit{single function} within parallelized scripts
is a unique phenomenon of \cryptojacking. Thus, in the second phase we
conduct a significantly prolonged run-time analysis of the candidate
sites, in which the sites receive \textit{no} external interaction and
hence should be idle after the initial rendering and set-up in
legitimate cases. However, if once the page is loaded, all JavaScript
is initialized, and the DOM is rendered, the CPU usage still remains on
a high level and the computation load is the result of repetitive
execution of a single function within the webpage's code base, we
conclude that the site hosts an \textit{active} \cryptojacking script.

\paragraph{Phase 3: Generalization of miner characteristics}

The run-time measurements of the first two phases limit our approach
to the detection of \textit{active} mining scripts. We thus might miss
mining sites that are inactive at the time of the test,
for instance due to programming errors in the site's JavaScript code,
a delayed start of the mining operation, or mining scripts waiting for
external events (\eg initial user interaction with the page).
To create a comprehensive overview, it is important to identify these
sites to document the \textit{intent of mining}.
To this end, we leverage the results of
the second phase to generate a set of static features of mining
scripts which we can be used to reevaluate the data from phase~1.

To this end, we first extract the JavaScript code from the validated
mining sites that is responsible for initiating and conducting the
mining operations. From this script code, we take both the URL and a
hash of its contents as two separate features. Furthermore, we collect
all parsed WebAssembly functions, sort them
and use the hash of the whole code base as the third
feature. We then apply each feature to our list of confirmed miners
from the previous phase and keep only those that describe at least a
certain number of miners. As the result, we obtain a set of
generalized fingerprints, which can identify common mining scripts
even in their \textit{inactive} state.  Applying these onto the data
collected during the first phase in combination with our list of
confirmed miners from the second phase yields to total number sites
with a web-based miner.

\subsection{Implementation}

We implemented the previously outlined approach into our custom web
crawler. In the following, we cover some details of the
implementation, which enable us to obtain accurate results on a large
scale in a real-world environment.

\subsubsection{Instrumented browser} 

We use a normal browser to visit all websites, in this case \textit{Google
Chrome}. This ensures that we will execute all scripts and support all modern
features needed to run a miner in the browser (see
\sect{sec:novelwebstandards}). By starting Chrome in headless mode, we can run
many instances simultaneously without the overhead of a GUI. Our crawler is
written in NodeJS and controls each instance via the \textit{DevTools
Protocol}~\citep{website:devtools}, which allows us to instrument the browser
and extract all necessary data. Amongst others, the protocol supports
retrieving all network traffic including WebSocket communication as well as
obtaining all parsed JavaScript and Wasm code by providing events like
\code{Debugger.scriptParsed} and \code{Network.webSocketFrameReceived}. As each
worker runs in a separate execution environment called \code{Target}, we need
to automatically attach to all targets related to the currently visited main
page in order to hook their events, too.

\subsubsection{Fake number of cores}

The number of logical cores of a
visitor's CPU is exposed in JavaScript via the
\code{hardwareConcurrency} property of the global \code{navigator}
object. This allows scripts to adjust the number of concurrent
WebWorkers according to the available hardware and is used by miners
to start the desired number of threads (usually one per core).
However, we do not want a single mining site to seize all available
resources on our server and interfere with simultaneous visits of
other websites. Furthermore, websites might employ checks on the
number of cores and not run if an unusually high number is observed,
thus preventing us from detecting them. Changing the returned value
can be achieved by overwriting the property as shown in
\fig{fig:hwconcurr}, without patching internals of the browser. By
using the DevTools Protocol's \code{addScriptToEvaluateOnNewDocument}
method, this script is injected into every page before any other
JavaScript execution starts and pretends that we only have \num{4}
logical cores, mimicking common desktop systems.

\begin{figure}[htbp]
\centering

\begin{minipage}[b]{0.95\linewidth}
\VerbatimInput[fontsize=\small, frame=single, numbers=left,
    numbersep=4pt]%
    {listings/hwconcurr.txt}
    \vspace{0mm}
\end{minipage}

\caption{Overwriting the native function to always return 4 instead of the actual number of cores.}
\label{fig:hwconcurr}
\end{figure}

\subsubsection{CPU Profiling}

Most importantly, instead of using standard Unix tools to measure the CPU load
on a \emph{per-process} level, we utilize the integered profiler of Chrome's
underlying JavaScript engine V8 to measure the load on a \emph{per-function}
level. We start this profiler, which is exposed by the DevTools Protocol via the
\code{Tracing} domain, with the \code{v8.cpu\_profiler.hires} option for more
accurate results. The profiler pauses the execution at a regular interval and
samples the call stack, which enables us to estimate the time spent in each
executed function. This way, we can not only determine if a \emph{single function}
consumes a considerable amount of CPU time, but also pinpoint the responsible
script in the website's code.

In order to achieve this, we aggregate the collected data for each unique call
stack. Wasm code itself cannot be profiled on a function level, so all samples
of it are just named \code{<WASM UNNAMED>}. However, from the call stack we can
still see how much time the Wasm code took and trace it back to the JavaScript
function which caused the call into Wasm in the first place (see
Table~\ref{table:profiling}). By comparing the time spent in a function with
the length of the profiling, we can estimate the caused CPU load for that
particular function. Note that if the same code is running in several workers
simultaneously, the combined time from all workers can be as high as the number
of cores times the length of the profiling, \eg our profiling for
\num{5000}~milliseconds with 4 CPU cores in phase~1 can result in a maximum time
of \num{20000}. Thus, taking the value of \num{14375} from
Table~\ref{table:profiling} as an example, would mean this function generated a
load of approximately \SI{72}{\%}.

\begin{table}[htbp]
  \centering\tablesize
  \begin{tabular}{
    l<{\hspace*{-4mm}}
    S[table-format=5.0]
    S[table-format=5.1]
  }
    \toprule
      \multicolumn{1}{l}{\bfseries Function name}
    & \multicolumn{1}{c}{\bfseries \# Samples}
    & \multicolumn{1}{c}{\bfseries Time in ms} \\
    \midrule
    \code{\small <WASM UNNAMED> }                       & 73938 & 14375.3 \\
    \code{\small Module.\_akki\_hash }                  &     1 &     0.1 \\
    \code{\small CryptonightWASMWrapper.hash }          &     4 &     0.6 \\
    \code{\small CryptonightWASMWrapper.workThrottled}  &    11 &     1.8 \\
    \code{\small (root) }                               &     0 &     0.0 \\
    \bottomrule
  \end{tabular}
  \caption{Example of a call stack with the aggregated amount of
      samples and time spent for each of its functions. }
  \label{table:profiling}
\end{table}

\subsection{Experimental Setup}

We used the aforementioned implementation to find instances of web-based
\cryptojacking in the wild. 
The following paragraphs briefly discuss
the key parameters of our experimental setup for each of the three
phases.

\paragraph{Phase 1} We conducted our study on the Alexa list of the
top 1 million most popular
sites\footnote{\url{http://s3.amazonaws.com/alexa-static/top-1m.csv.zip}}.
We visit the front page of each site and wait until the browser fires
the load event or a maximum of \num{30} seconds pass. Furthermore,
to allow for sites that dynamically load further content, we wait an
additional \num{3} seconds or until no more network requests are
pending. We then start the CPU profiler and measure all code execution
for \num{5} seconds and flag the site as suspicious, if there is a function
with more than~\perc{5} load on average. As the most CPU-heavy function on each
website of the Alexa Top 1 million had an average load of only $\perc{0.2} \pm
\num{3.21}$, we reckon that a value of 5\% or more warrants further
investigation. We also flag the site for extendend analysis, if either any Wasm
code or more than \num{3} workers are used, which is equal or more than all the
CPU cores we pretend to have. For this phase, we used a single server with
\num{24} CPU cores and \num{32} GB of RAM running \num{24} simultaneous
crawlers backed by Chrome v67.0.3396 over a time span of \num{4} days.

\paragraph{Phase 2} The detailed verification of suspicious sites uses
the same general setup as the first phase. However, here we only run
one crawler on a smaller server with \num{8} CPU cores. By visiting
the websites one-by-one and profiling for a longer time of \num{30}
seconds, we can more accurately determine if a website contains a
mining script. If there is one function in the code base that results
in an average load of~\perc{10} or more, we label it as a confirmed
and active miner. We argue that while a value lower than 10\% certainly would
make the miner very hard to detect, it also severly thwarts the ability to make
money with cryptojacking. Furthermore, such slow mining does not even seem to
be supported by popular mining scripts, as we will describe shortly.

\paragraph{Phase 3} In the final step, we create the fingerprints as
outlined in \sect{sec:general-approach} using the code of the
confirmed miners. However, we only keep the fingerprints shared by at
least \perc{1}~of all miners. This restrictive measure ensures
that only mining scripts with multiple validated instances produce 
fingerprints and, thus, avoids accidental inflation of potential phase 2
classification mistakes. The resulting fingerprints are then applied to the
collected data from the first phase, yielding the final number of websites
employing \cryptojacking on their visitors.

To validate that our implementation and setup are working as intended, we created a testbed with the two popular implementations that start without the user's consent: \coinhive and CryptoLoot.
This testbed consists of 24 locally hosted pages, which each contain one of the miners at a different throttling levels between 0\% and 99\%.
Interestingly, even if the miner is configured with a throttle as high as 99\%, so that it should utilize only 1\% CPU, we can confirm it as a miner with our 10\% threshold.
Looking into the implementation of the throttling, we find that the code never sleeps for longer than two seconds between hash calculations, which makes it impossible to actually use very low throttling values.
We also confirm this by monitoring Chrome's CPU usage with \code{htop} and find that no matter how high we set the throttling, the load on our machine never drops to below 20\%.
As our implementation is able to successfully detect all miners in the testbed, regardless of the used throttling value, we are confident its ability to find active cryptojacking scripts.

\subsection{Prevalence}
\label{sec:prevalence}

As first result of our crawling, we identify \num{4627} suspicious
sites in the Alexa ranking using the methodology and parameters
outlined in the previous sections.  Out of these, \num{3028} are
flagged for having a load-intensive function, \num{3561} for using at
least as many workers as CPU cores we pretend to have and \num{2477}
for using Wasm. Note that these sets overlap, as for example
the usage of Wasm often implies a CPU intensive application.

The detailed analysis of these \num{4627} suspicious sites results in
\num{1939} sites with a continuously high CPU usage over a profiling
for \num{30} seconds. We use the resulting set of scripts 
for the third phase to build
fingerprints of the most popular miners, resulting in \num{15}
hashes of JavaScript code, \num{12} hashes of Wasm code bases, and
\num{8} script URLs. The latter can be found in
Table~\ref{table:fingerprint-urls}. After applying these fingerprints,
we obtain the final number of \num{2506} websites, which are very
likely to employ \cryptojacking. Table~\ref{table:prevalence}
summarizes our results.

\begin{table}[htbp]
  \centering\tablesize
  \begin{tabular}{
    S[table-format=1.0]
    l
    S[table-format=4.0]
    S[table-format=1.2]
  }
    \toprule
    \multicolumn{1}{c}{\bfseries Phase} &
    \multicolumn{1}{l}{\bfseries Result} &
    \multicolumn{1}{r}{\bfseries \# Websites} &
    \multicolumn{1}{r}{\bfseries \% of Alexa} \\
    \midrule
    1 & Suspicious sites            & 4627 & 0.46 \\
    2 & Active \cryptojacking sites & 1939 & 0.19 \\
    3 & Total \cryptojacking sites  & 2506 & 0.25 \\
    \bottomrule
  \end{tabular}
  \caption{Prevalence of web-based miners in the Alexa Top~1 million websites.}
  \label{table:prevalence}
\end{table}

\begin{table}[htbp]
  \centering\tablesize
  \begin{tabular}{
    l
    S[table-format=3.0]
  }
    \toprule
    \multicolumn{1}{l}{\bfseries URL} &
    \multicolumn{1}{r}{\bfseries \# Occurrences} \\
    \midrule
    \url{ //coinhive.com/lib/coinhive.min.js                } & 656 \\
    \url{ //advisorstat.space/js/algorithms/advisor.wasm.js } & 311 \\
    \url{ //www.weather.gr/scripts/ayh9.js                  } &  68 \\
    \url{ //aster18cdn.nl/bootstrap.min.js                  } &  59 \\
    \url{ //cryptaloot.pro/lib/crypta.js                    } &  46 \\
    \url{ //gninimorenom.fi/sytytystulppa.js                } &  35 \\
    \url{ //coinpot.co/js/mine                              } &  27 \\
    \url{ //mepirtedic.com/amo.js                           } &  22 \\
    \bottomrule
  \end{tabular}
  \caption{Common script URLs responsible for the creation of the
      mining workers, which resulted in fingerprints}
  \label{table:fingerprint-urls}
\end{table}

During manual investigation of a sample of the additional \num{567}
sites only detected in phase~3, we found five reasons why our dynamic
analysis missed these miners:
(1)~A script for web-based mining is included, but the miner is never
started. (2)~The miner only starts once the user interacts with the web
page or after a certain delay. (3)~The miner is broken---either because
of invalid modifications or because the remote API has changed (as it
was the case for \coinhive earlier this year). (4)~The WebSocket backend
is not responding, which prevents the miner from running. (5)~The miner
is only present during some visits, \eg to hinder detection or due to ad
banner rotation

This analysis confirms the need for a three-step identification process,
as only the combination of phase~2 and~3 enable us to determine a
comprehensive picture of current \cryptojacking in the websites of the
Alexa ranking.

\subsection{Effectiveness of Countermeasures}
\label{sec:blacklists}

To both compare our findings to existing approaches for the detection of
\cryptojacking and to validate our results, we select three popular solutions
to block miners in the browser. For one, we use the \textit{NoCoin adblock
  list}\footnote{\url{https://github.com/hoshsadiq/adblock-nocoin-list/}},
which is a generic list for adblockers, such as Adblock Plus or uBlock
Origin and is now also used by Opera's built in adblocker. For the
remainder of this section, we refer to this list as
\textit{Adblocker}. Furthermore, we include the blacklists used by the
two most popular Chrome extensions with the purpose of blocking
web-based miners: \textit{No
  Coin}\footnote{\url{https://chrome.google.com/webstore/detail/gojamcfopckidlocpkbelmpjcgmbgjcl}}
with \num{566692} users and
\textit{MinerBlock}\footnote{\url{https://chrome.google.com/webstore/detail/emikbbbebcdfohonlaifafnoanocnebl}}
with \num{161630} users. We extract the detection rules these extensions
contain and translate them into SQL statements while preserving the wildcards,
in order to apply them to the data collected during our crawl of the Alexa Top
1 million sites. The number of identified miners for each system are presented
in Table~\ref{table:blacklists} in the first column. The other columns of this
table compare these results for each system with the \num{2506} websites we
identified as miners. The second column reports on the
intersection of both lists, that is the number of sites on which both
approaches are in agreement. Accordingly, the last two columns each contain
the number of sites that one approach reported, but not the other.

\begin{table}[htbp]
  \centering\tablesize
  \begin{tabular}{
    l
    S[table-format=4.0]
    S[table-format=4.0]
    S[table-format=3.0]
    S[table-format=4.0]
  }
    \toprule
      \multicolumn{1}{l}{\bfseries Blacklist}
    & \multicolumn{1}{c}{\bfseries \# Detections}
    & \multicolumn{1}{c}{\bfseries \# Both}
    & \multicolumn{1}{c}{\bfseries \# Only they}
    & \multicolumn{1}{c}{\bfseries \# Only we} \\
    \midrule
    { Minerblock } & 1599 & 1402 & 197 & 1104 \\
    { No Coin }    & 1217 & 1039 & 178 & 1467 \\
    { Adblocker }  & 1136 & 1049 &  87 & 1457 \\
    \bottomrule
  \end{tabular}
  \caption{Detection results of our approach and three common
    blacklists as absolute numbers.}
  \label{table:blacklists}
\end{table}

Unsurpisingly, our approach mixing static and dynamic analysis clearly outperforms
the three static blacklists and spots a considerable amount of additional
web-based miners. Moreover, the large overlap in sites that both we and the
extensions found, validates that our approach and shows that it is indeed
suitable to detect \cryptojacking in the wild.

There are, however, a few sites that our approach
misses, but the blacklists detect. Manual analysis of a subset showed that
besides overly zealous lists, the main reason is that we can only learn
fingerprints of \textit{active} miners. For example, some website owners
copied CoinHive's script to host it on their own servers a few months ago.
Meanwhile, all these mining scripts stopped working, as \coinhive changed its
API used in the communication with the pool. Therefore, while this probably
represents a cluster of inactive miners, we are unable to detect them, as no
fingerprint for any of the scripts could be generated in the third phase, due
to the fact that the \textit{whole} cluster was inactive at the time of
analysis.

The existing blacklists on the other hand can detect them, as their
rules are curated by humans, which allows them to apply a couple of generic measures. 
For example, most blacklists include a rule
for \code{*/coinhive.min.js}. In contrast, our static indicators are generated in a fully 
automated fashion, based on code characteristics from dynamically validated miner instances.  
In this process, we cannot 
generalize our list of fingerprinted full script URLs towards partial URLs or even only 
filenames without manual review, as this could lead to misclassifications.  
For instance, in our dataset such an attempt would end up in all scripts named 
\code{*/bootstrap.min.js} being blacklisted because a widespread mining script uses this
benign-sounding name (see Table~\ref{table:fingerprint-urls}).

\subsection{Distribution}
\label{sec:distribution}

\begin{figure}[htbp]
  \centering
  \includegraphics[width=0.8\columnwidth]{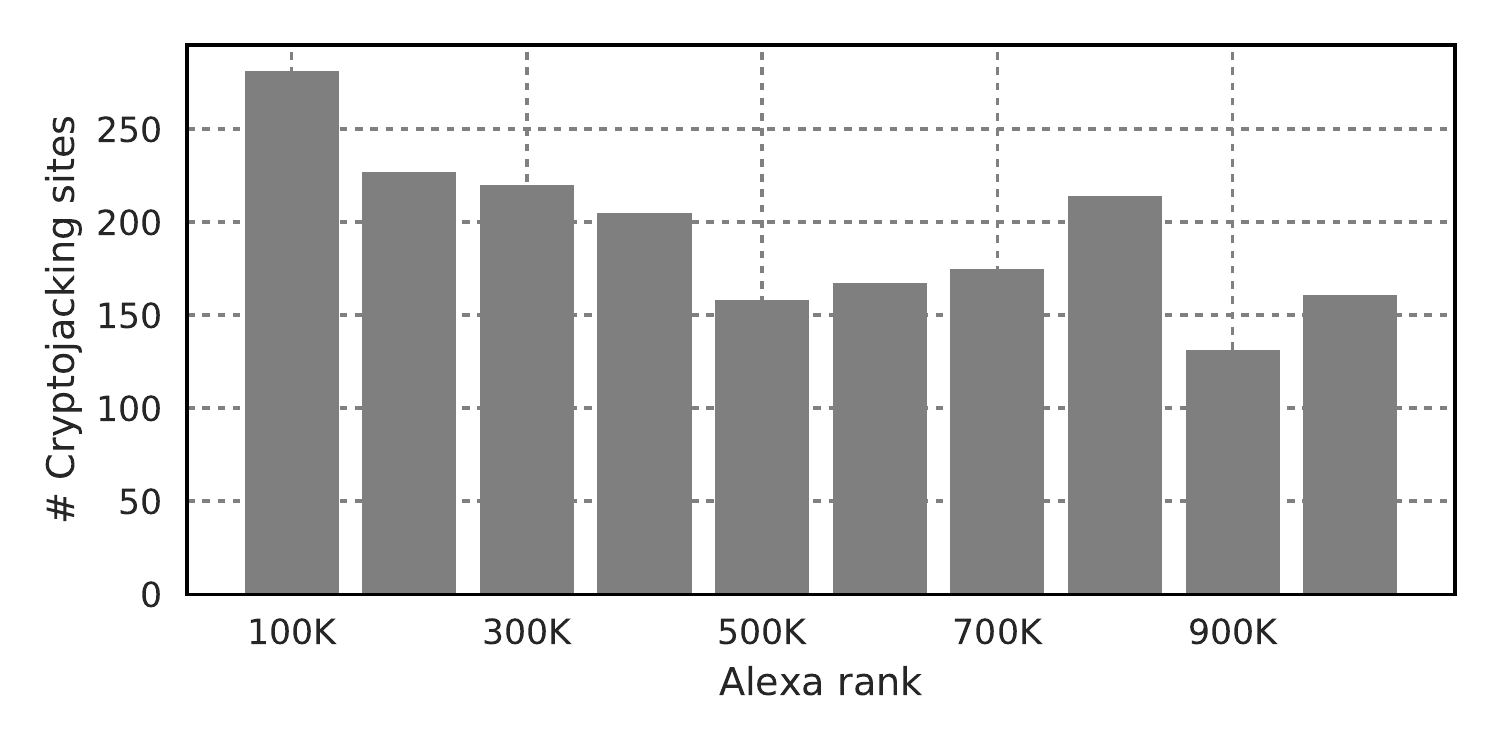}
  \vspace*{-2mm}
  \caption{Distribution of \cryptojacking by popularity using the
    Alexa ranking.}
  \label{fig:dist-alexa}
\end{figure}

Next, we investigate what evidence can be drawn on sites that host
\cryptojacking scripts and if their usage is dominant on certain parts
of the Internet or if their prevalence is more uniformly
distributed. First of all, we check if there is a connection between
the popularity of a website and its likelihood to use
\cryptojacking. For this, we use the rank provided by Alexa and
included in our initial selection of the top~\num{1} millions sites. A low
number indicates a highly popular site, \eg\xspace \code{google.com} has the
rank \num{1}. While there is a slight trend towards the lower ranks,
\cryptojacking is indeed a wide-spread phenomenon not limited to
popular websites (see Figure~\ref{fig:dist-alexa}).

Furthermore, we want to know where \cryptojacking sites operate and
therefore examine the location of the hosting servers, by resolving the
website's IP address with our local DNS server and using a free
geolocation
database\footnote{\url{https://dev.maxmind.com/geoip/geoip2/geolite2/}}.
We find that the majority of \cryptojacking sites are hosted in the
United States, followed by Russia and Germany (see~\tab{table:dist-countries}).

\begin{table}[htbp]
  \centering\tablesize
  \begin{tabular}{
    l
    S[table-format=4.0]
    l
    S[table-format=4.0]
  }
    \toprule
    \multicolumn{1}{l}{\bfseries Hosting country} &
    \multicolumn{1}{c}{\bfseries \# Sites} &
    \multicolumn{1}{l}{\bfseries Hosting country} &
    \multicolumn{1}{c}{\bfseries \# Sites} \\
    \cmidrule(rl){1-2}\cmidrule(rl){3-4}
    United States & 1118 & France        &  124 \\
    Russia        &  508 & Netherlands   &  111 \\
    Germany       &  203 & Others        &  442 \\
    \bottomrule
  \end{tabular}
  \caption{Five most common countries which host \cryptojacking
    miners in websites of the Alexa ranking.}
  \label{table:dist-countries}
\end{table}

Finally, we check what content these sites provide in order to attract
visitors. For this, we use data from Symantec's WebPulse Site
Review\footnote{\url{http://sitereview.bluecoat.com/lookup}}.
Table~\ref{table:dist-categories} lists the most popular categories and shows
that websites offering entertainment in general or pornography in particular
are most common, with the topics technology and business in the follow-up.
Furthermore, \num{584} sites are labeled as suspicious and another \num{541} are
labeled as malicious by Symantec.

\begin{table}[htbp]
  \centering\tablesize
  \begin{tabular}{
        l<{\hspace*{3mm}}
        S[table-format=3.0]
      }
    \toprule
    \multicolumn{1}{l}{\bfseries Category} &
    \multicolumn{1}{c}{\bfseries \# Sites} \\
    \midrule
    Entertainment                 & 237 \\
    Malicious Sources/Malnets     & 231 \\
    Pornography	                  & 184 \\
    Technology/Internet           & 138 \\
    Business/Economy              & 103 \\
    Education                     &  99 \\
    Piracy/Copyright Concerns     &  96 \\
    News/Media                    &  95 \\
    Games                         &  80 \\
    TV/Video Streams              &  63 \\
    \bottomrule
  \end{tabular}
  \caption{Ten most common categories for websites with \cryptojacking.
  One website can have multiple categories.}
  \label{table:dist-categories}
\end{table}

\section{Analysis of \cryptojacking}
\label{sec:analysis}

After identifying web-based miners in the Alexa ranking, we proceed to
analyze the efficacy of these miners in detail. For this analysis, we
focus on all \emph{active miners} discovered in phase 2 of our study, that
is, websites that immediately begin mining when visited by a
browser. We start by estimating the profit of these \num{1939}~mining
websites and answer the question of whether such mining can generate
significant income (Section~\ref{sec:revenue}).
We then investigate how aggressively these web pages stress the
visitors' CPUs to shed light on the stealthiness of current
\cryptojacking in the web (\sect{sec:greediness}).
Finally, we determine how many different implementations of miners
exist in our dataset and whether these differ due to customization and
obfuscation (\sect{sec:clustering}).

\subsection{Revenue Estimation}
\label{sec:revenue}

Determining the exact revenue of web-based miners is a non-trivial
task, as the profit depends on several factors, such as the popularity
of a website, its content, the visitor's hardware as well as the
current price of the cryptocurrency. Consequently, we estimate the
revenue at different levels of granularity. First, we determine a
rough upper bound by calculating the potential revenue obtained when
mining on a very large video-streaming website.
Then, we dissect the code of the active miners in our study and
extract \coinhive site-keys and wallet addresses that enable us to
identify the cryptocurrencies and estimate the profit obtained by
individual miners.

\subsubsection{Upper-bound estimate}
\label{sec:revenue:example}

\newcommand{\pornhub}{Pornhub\xspace}

Video streaming and multimedia websites naturally attract the
attention of visitors for a longer time than other content on the
web. For a first estimate of the potential profit that a large website
may generate, we consider the \pornhub video-streaming platform for
adult content, for which detailed statistics about visits are
published~\citep{website:pornhubstats}.
In 2017, for instance, the website attracted a total of
\num{81}~million visitors per day. Each of these visitors spent
roughly \SI{10}{minutes} (\SI{9}{minutes} \SI{59}{seconds}) on the
website, which sums up to \num{13.5}~million hours spent on \pornhub
world-wide each day. This offers a huge opportunity for advertisement
and web-based mining likewise.\\

\newcommand{\registered}{\textregistered\xspace}
\newcommand{\trademark}{\texttrademark\xspace}

\newcommand{\intelxeon}{Intel\registered Xeon\registered}
\newcommand{\intelcore}{Intel\registered Core\trademark}

\begin{table}[htbp]
  \centering\tablesize
  \begin{tabular}{
    l<{\hspace*{-4mm}}
    S[table-format=2.0,table-space-text-post={\,MB}]<{\,MB}
    S[table-format=2.1]
    S[table-format=3.1]
  }
    \toprule
      \multicolumn{1}{l}{\bfseries CPU model}
    & \multicolumn{1}{r}{\bfseries Cache size}
    & \multicolumn{2}{c}{\bfseries Hashes/s} \\
    \cmidrule(rl){3-4}
      \multicolumn{1}{l}{Product name and clock speed}
    & \multicolumn{1}{c}{L2/L3}
    & \multicolumn{1}{c}{Core}
    & \multicolumn{1}{c}{CPU} \\
    \midrule
    { \intelxeon E5-1650 v3 @ \SI{3.50}{\GHz} } & 15 & 22.2 & 148.9 \\ %
    { \intelcore i7-7700K @ \SI{4.20}{\GHz} }   &  8 & 21.4 & 115.3 \\ %
    { \intelcore i7-6820HQ @ \SI{2.70}{\GHz} }  &  8 & 23.2 & 90.2\\ %

    { \intelcore i7-5557U @ \SI{3.10}{\GHz} }   &  4 & 21.1 &  35.5 \\ %
    { Apple A11 Bionic APL1W72 }                &  8 & 16.0 &  35.1 \\ %
    { HiSilicon Kirin 620 @ \SI{1.20}{\GHz} }   &  2 &  2.0 &  11.6 \\ %
    \bottomrule
  \end{tabular} \\ %
  \caption{Performance of different CPUs with \coinhive.}
  \label{table:cpus}
\end{table}

The actual revenue, however, depends on the specific computational
power of the visitors' hardware and the implementation of the
miner. For our analysis, we thus focus on the \coinhive library and
measure its run-time performance for different desktop and mobile
CPUs. Results of this experiment are shown in \tab{table:cpus}, where
the performance is presented in hashes per second for one core and the
entire CPU. Due to the memory-bound proof-of-work function, the hash
rate varies only slightly between the different CPU models when
executed on one core. The only exception is the HiSilicon CPU whose
cache is limited to \SI{2}{\mega\byte} and thus is not suitable for
computing the CrypoNight hash.

If we assume a rate of \hashrate for an average CPU, we arrive at a
revenue of \phrevenue~per day for the entire \pornhub platform, under
CoinHive's payout ratio\footnote{\label{payout}\coinhive's payout
  ratio at the time of writing: \chpayout.}. That is, a miner could
potentially earn \SI{50208}{USD} per day on \pornhub for the
current exchange rate\footnote{According to \url{https://coinmarketcap.com/},
at time of our measurement crawl in May 2018} of \SI{1}{XMR} =
\SI{225}{USD}.

\subsubsection{Expected per-site revenue}
\label{sec:revenue:expected}

We proceed to estimate the expected revenue for the active miners
identified in \sect{sec:identification}. In particular, we make use of
the \similarweb service to quantify the number of visits as well as
the average duration for the websites hosting the miners. The results
of this analysis are shown in \tab{table:visitingstats}, where we
include the \num{10}~most profitable sites identified during our analysis.
These sites are able to generate between \num{0.53}~and~\SI{1.51}{XMR} per day,
that is, \num{119} to \SI{340}{USD}.  Given that the revenue is achieved
without the consent of the visitors and visual indications, this is still a
notable profit. However, we conclude that current cryptojacking is not as
profitable as one might expect and the overall revenue is moderate.

\begin{table}[htbp]
  \centering\tablesize
  \begin{tabular}{
    l<{\hspace*{-5mm}}
    S[table-format=1.1,table-space-text-post={\,M}]<{\,M}
    S[table-format=2.0,table-space-text-post={'00''}]
    S[table-format=2.0,table-space-text-post={\,K}]<{\,K}
    S[table-format=2.1]
  }
    \toprule
    & \multicolumn{1}{c}{\bfseries Visitors}
    & \multicolumn{1}{c}{\bfseries Duration}
    & \multicolumn{1}{c}{\bfseries Core hours}
    & \multicolumn{1}{c}{\bfseries Revenue*} \\
    \cmidrule(rl){2-2}
    \cmidrule(rl){3-3}
    \cmidrule(rl){4-4}
    \cmidrule(rl){5-5}
    & \multicolumn{1}{c}{per day}
    & \multicolumn{1}{c}{per visit}
    & \multicolumn{1}{c}{per day}
    & \multicolumn{1}{c}{XMR per day} \\
    \midrule
    { cinecalidad.to }     &   1.3 &  4'10'' &    89 &   1.5\\
    { mejortorrent.com }   &   0.8 &  4'30'' &    60 &   1.1\\
    { kinokrad.co }        &   1.3 &  2'29'' &    54 &   1.0\\
    { ianimes.co }         &   0.2 & 13'07'' &    39 &   0.7\\
    { india.com }          &   1.3 &  1'27'' &    32 &   0.6\\
    { ddmix.net }          &   0.4 &  5'06'' &    38 &   0.6\\
    { seriesypelis24.com } &   0.4 &  5'22'' &    35 &   0.6\\
    { seriesblanco.com }   &   0.4 &  6'06'' &    36 &   0.6\\
    { ekinomaniak.tv }     &   0.2 & 10'10'' &    33 &   0.6\\
    { kickass.cd }         &   0.3 &  5'24'' &    30 &   0.5\\

    \bottomrule
  \end{tabular}
  \caption{Visiting statistics for the top-\num{10} sites containing miners.}
  \label{table:visitingstats}

  \vspace{1ex}
  \centering *~Estimated based on \coinhive's payout ratio\footnoteref{payout} and \hashrate.
\end{table}

\newcounter{boxplot}
\setcounter{boxplot}{2}

\begin{figure}[htbp]
\ifnum\the\value{boxplot}=0
  \centering
  \includegraphics[width=\columnwidth]{figures/visits-box.pdf}
  \caption{Distribution of visits to web pages identified to use
  web-based miners and the duration per visit.}
\fi
\ifnum\the\value{boxplot}=1
  \centering \subfigure[Visits per day]{
    \includegraphics[width=0.9\columnwidth, trim={0 5mm 0 0}]{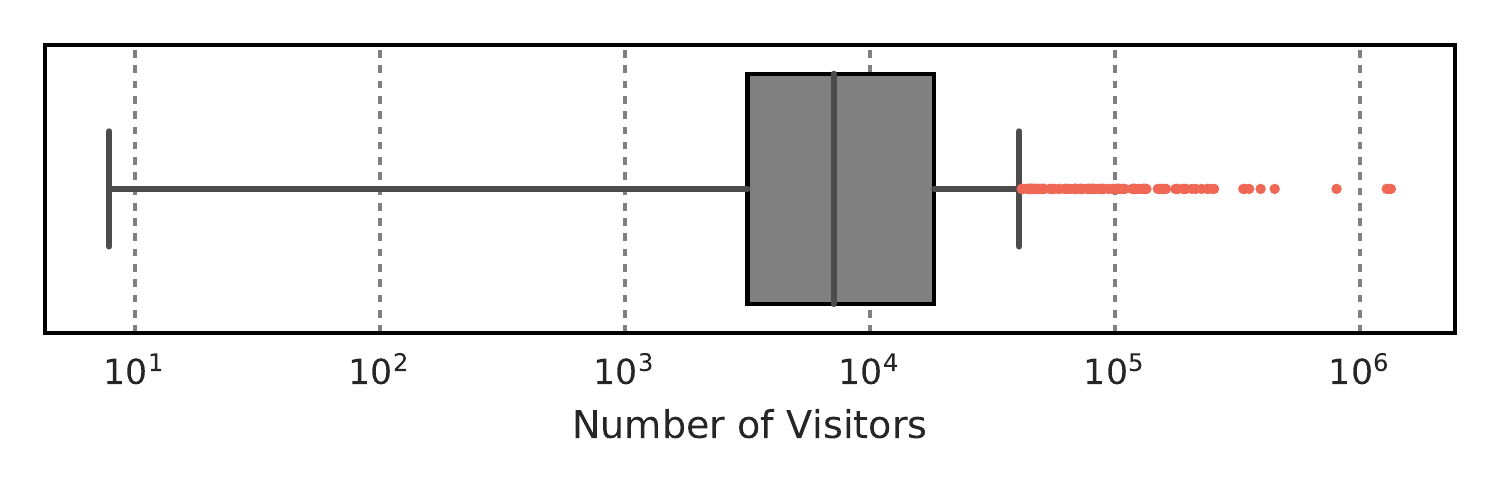}}
  \centering \subfigure[Duration per visit]{
    \includegraphics[width=0.9\columnwidth, trim={0 5mm 0 3mm}]{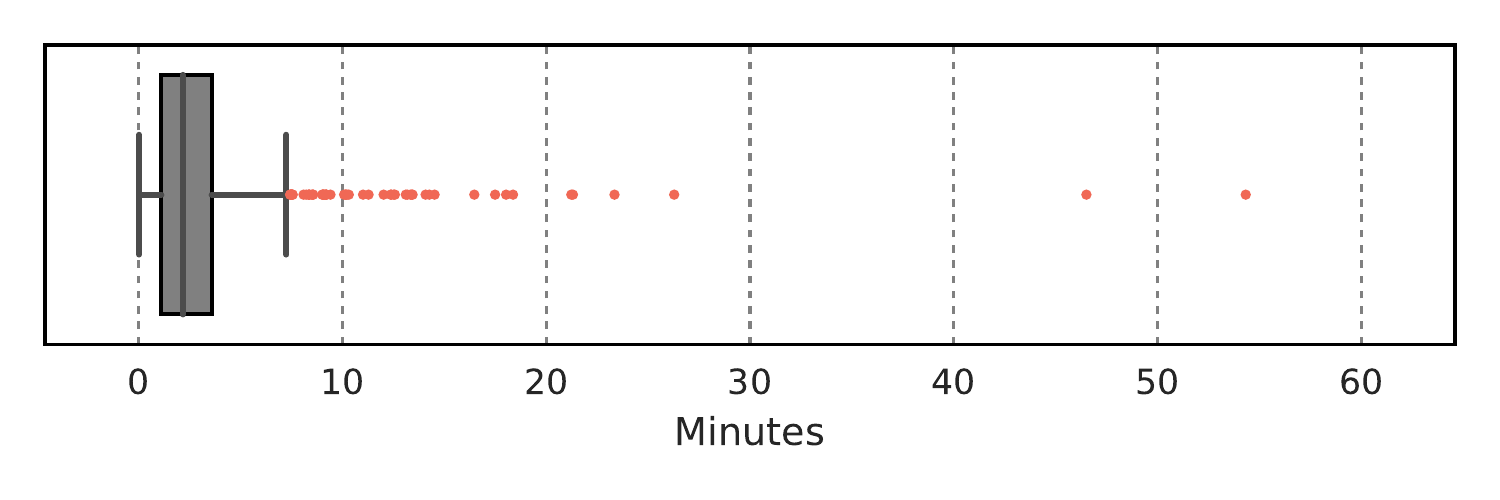}}
  \vspace{-3mm}
  \caption{Distribution of visits to web pages identified to use
  web-based miners and the duration per visit.}
\fi
\ifnum\the\value{boxplot}=2
  \centering
  \includegraphics[width=0.80\columnwidth, trim={0 0 0 3mm}]{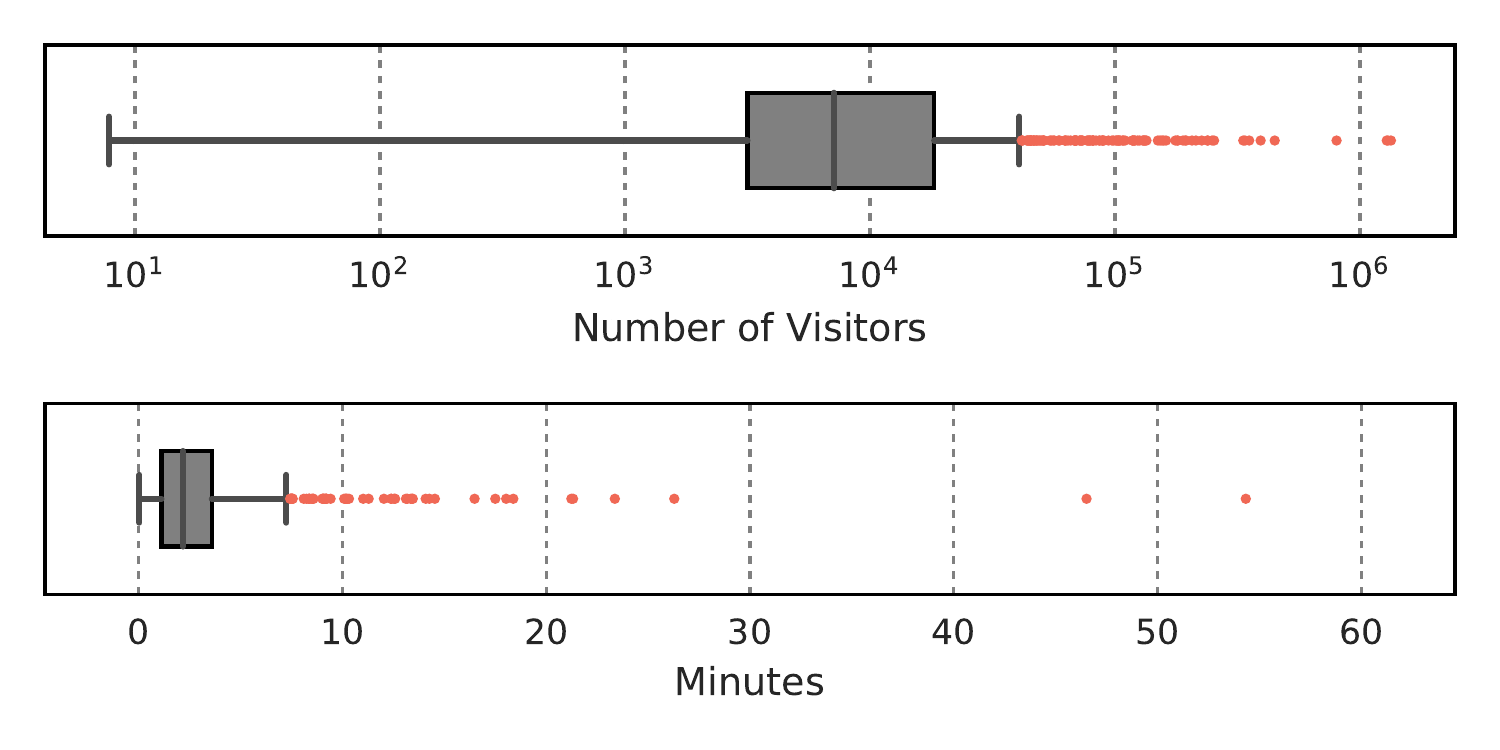}\vspace{-3mm}
  \caption{Distribution of visits to web pages identified to use
  web-based miners (top) and the duration per visit (bottom).}
\fi
  \label{fig:visitingdistr}
\end{figure}

\subsubsection{Average revenue}

We continue to asses the average profit made by mining on the web. To
this end, we inspect the distribution of visits per day and the
average duration of these in \fig{fig:visitingdistr}.
The websites with the largest outreach in our dataset
(\code{cinecalidad.to}) has \num{1.3}~million visits. A different site
(\code{ianimes.co}) attracts less visitors, but engages them to stay
\num{13}~minutes on the web page. On average these websites attract
\num{24721}~visitors per day and keep them for roughly \SI{3}{minutes}
on average.
Overall, we thus observe a range of \num{0.17}~to~\num{89000}\,core
hours, with a mean of \num{1550}\,core hours. With a hash rate
of~\hashrate and \coinhive's payout ratio\footnoteref{payout}, a miner
earns about \empirical{\SI{5.8}{USD}} per day and website on average,
which supports our observation that web-based \cryptojacking currently
provides only limited profit.

Next, we group websites that make use of the same site-key to
calculate the overall revenue of mining entities.
Tracking these relations provides valuable insights on the landscape of
\cryptojacking, as we can identify attackers that deploy miners on
multiple websites.
Note, that this analysis is not limited to \coinhive, but applies to
any variant using the original implementation. We thus observe a large
variety of unique site identifiers. A few instances make use of
nondescript values such as \code{X} or \code{abc}, though, which we
filter out for this particular measurement.
\fig{fig:keygroups} depicts the frequency of websites per site-key in
bins of \num{5}.

\iftrue
\begin{figure}[htbp]
  \centering
  \includegraphics[width=0.8\columnwidth]{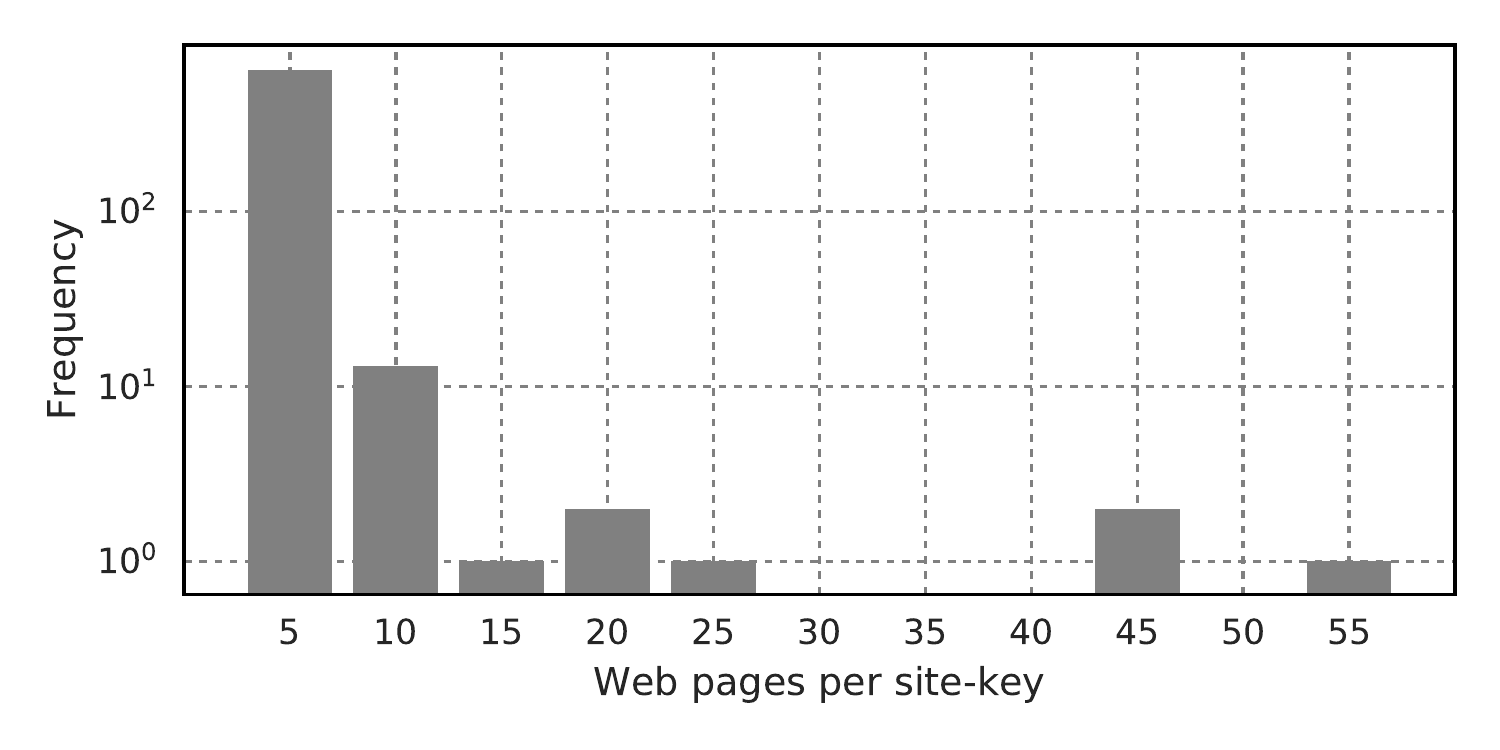}%
  \vspace*{-3mm}%
  \caption{Distribution of websites per site-key.}
  \label{fig:keygroups}
\end{figure}

\else
\begin{figure}[htbp]
  \centering
  \includegraphics[width=\columnwidth]{figures/keys-distr.pdf}%
  \vspace*{-3mm}%
  \caption{Websites using multiple \coinhive site-keys.}
  \label{fig:keygroups}
\end{figure}
\fi

Our analysis shows that site-keys are indeed reused across different
websites.  Few \cryptojackers even pool forces across up to
\num{40}~to~\num{55}~websites, while the majority of attackers appears
to act on their own or, more likely, utilize \coinhive's function
of aggregating multiple site-keys with one account.
Moreover, we observe that \num{23}~web pages make use of at least
two site-keys. These websites thus connect clusters of miners.

\subsubsection{Used cryptocurrencies}
\label{sec:revenue:coins}

Several cryptocurrencies, such as Monero, Electroneum, and ByteCoin,
rely on the CryptoNight proof-of-work function and thus can be mined
using the same implementation with only minor modifications.
Consequently, the community has started to repurpose the original code
of \coinhive to mine alternative currencies, but also to use
alternative pools with less severe payout fees.
Some of these implementations require specifying the wallet address
that is used to track the revenue in mining pools. As a side effect,
this address can be observed within the WebSocket communication of the
miner. \fig{fig:websocket_comm} shows a captured message that is used
to communicate with a proxy, where the value of the \code{login}
field represents the Monero wallet to mine for.

\begin{figure}[htbp]
\centering
\begin{minipage}[b]{0.85\linewidth}
\begin{Verbatim}[fontsize=\small, frame=single, numbers=left, numbersep=4pt]
  "identifier":"handshake",
  "pool":"supportxmr.com",
  "login":"4676xXzU5tXfx4tDdDS...WivxG9c1ih6V",
  "password":"",
  "userid":"",
  "version":4
\end{Verbatim}
\end{minipage}
\caption{Mining communication over WebSockets.}
\label{fig:websocket_comm}
\end{figure}

Wallet addresses follow a very strict pattern and normally contain a
checksum. In consequence, these addresses can be easily identified and
located in network traffic. A Monero wallet, for instance, is a base58
encoded binary string, where the last four~bytes are the Keccak hash
(as submitted to the SHA-3 contest, but not the FIPS~\num{202}
compliant implementation) of the preceding bytes. While the wallet
address encodes more information, necessary for conducting
transactions, this information is irrelevant for locating the address
in network communication.

We use this strategy to identify wallet addresses in the traffic of the
discovered active miners in our dataset. \tab{table:wallets} summarizes
our findings.
We identify a total of \num{36}~different wallets, with a large
majority accounting for the cryptocurrency Monero, while for
Electroneum and Bytecoin we find only \num{3}~and~\num{2}~addresses,
respectively.  Moreover, we spot two instances of lesser-known
currencies, Intense Coin and Graffiti Coin. These results reflect the
popularity and market capitalization of the individual currencies.

Additionally, we are able to identify \num{570}~site-keys corresponding
to the \coinhive service, which at the time being also mines Monero
coins. Due to \coinhive's separation of site-keys and wallets, the
service, however, is not bound to a particular cryptocurrency and may
change to or incorporate other markets in the future. We thus consider
\coinhive a service without specific currency.

\begin{table}[htbp]
  \centering
  {\vspace{-2mm}
  \begin{tikzpicture}
  \node (dummy) {};
  \node (pie) at ($(dummy.east)+(5.8,0.2)$)
  {\includegraphics[height=2.5cm]{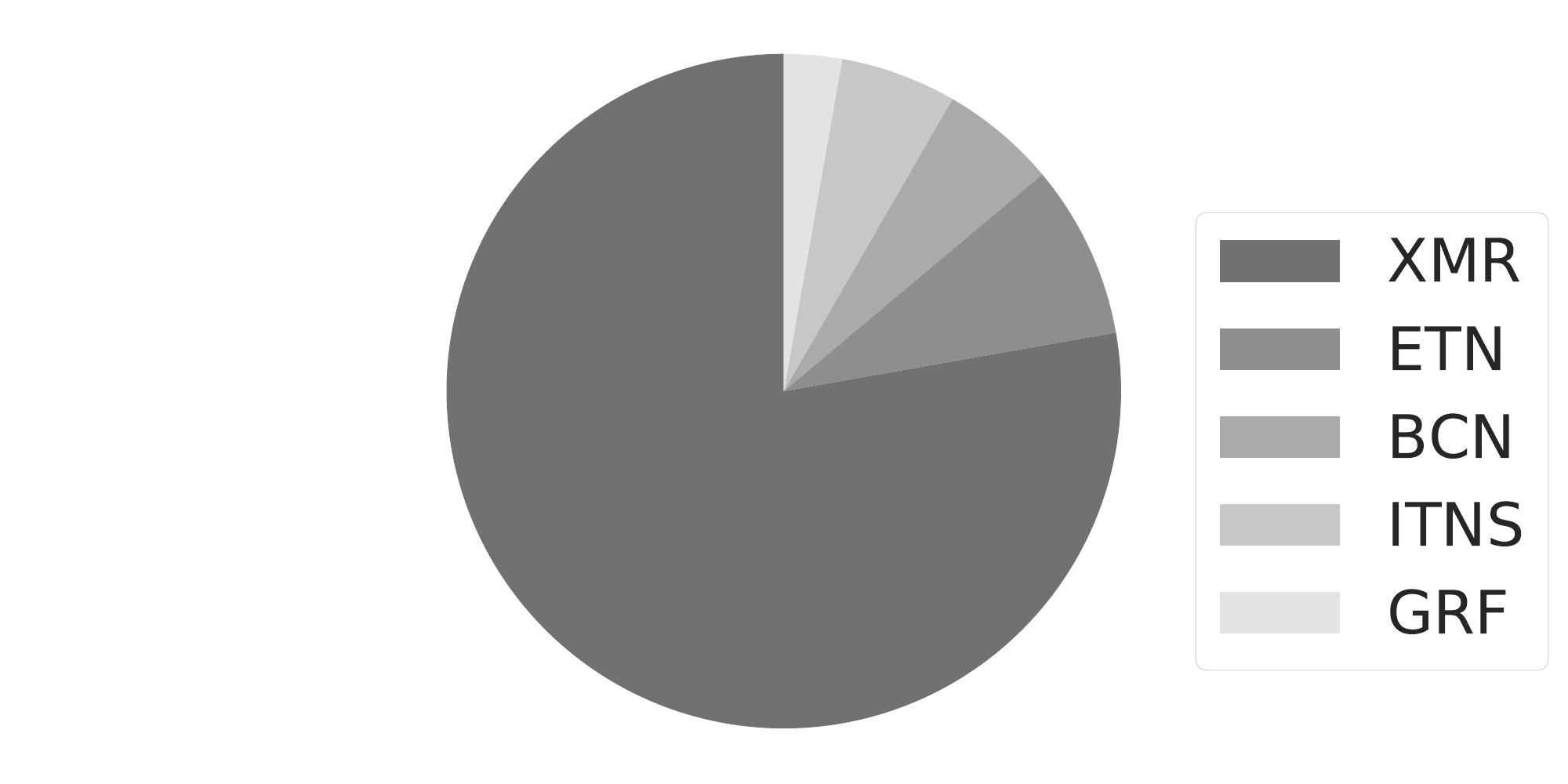}};
  \node (tab) at ($(dummy.east)+(2.0,0.0)$)
  {\small\begin{tabular}{
    l
    S[table-format=3.0]
  }
    \toprule
      \multicolumn{1}{l}{\bfseries Crypto currency}
    & \multicolumn{1}{r}{\bfseries \#} \\
    \midrule
    { Monero (XMR)        } &  28 \\
    { Electroneum (ETN)   } &   3 \\
    { Bytecoin (BCN)      } &   2 \\
    { Intense Coin (ITNS) } &   2 \\
    { Graffiti Coin (GRF) } &   1 \\
    \midrule
    { \coinhive           } & 570 \\
    \bottomrule
  \end{tabular}
  };
\end{tikzpicture}
  }
  \caption{Cryptocurrency wallets used for mining.}
  \label{table:wallets}
\end{table}

\subsubsection{Quantitative estimate}

Finally, we determine the profit associated with the wallets
identified in the previous section. Unfortunately, tracking cash-flows
for cryptocurrencies based on the CryptoNote protocol is only possible
in a few, rare scenarios and under very specific
preconditions~\citep[see][]{MoeSosHei+18, KumCleFisTopSax17}.
We thus have to revert to a more abstract way of mapping payouts to
miners. In particular, we query mining pools for the wallet addresses
we have found and use the balance maintained by the pool for each
miner.  This enables us to determine a lower bound of the profit made,
despite the anonymity guarantees of the CryptNote protocol.

With this procedure, we can track down \SI{15000}{USD} worth of
cryptocurrencies, as of the day of writing, across different wallet
addresses. In line with the breakdown in \tab{table:wallets}, Monero
dominates the measure with \perc{99.4} of the value.

\subsection{Greediness vs. Stealthiness}
\label{sec:greediness}

The revenue of a \cryptojacking campaign may vary a lot, depending on how
aggressive the miner occupies the visitor's CPU cores. Using large amounts of
processing power earns the most money, but simultaneously may raise
suspicion due to an unresponsive computer and audible fan noise. An attacker
thus has to strike a balance between profit and stealthiness in practice.

Many popular implementations of web-based miners allow for the
configuration of a throttling value. While \coinhive's default value
is \num{0}, their example recommends a value of \num{0.3}, which means
the miner only uses \perc{70}~of the available computing power by
constantly monitoring the current and maximum possible hash rate and
idling or mining accordingly. The data we have gathered in \emph{Phase~2}
for the validation of miners allows us to approximate the CPU load and
thus also the throttling value chosen by the website operator.

\begin{figure}[htbp]
  \centering
  \includegraphics[width=0.80\columnwidth]{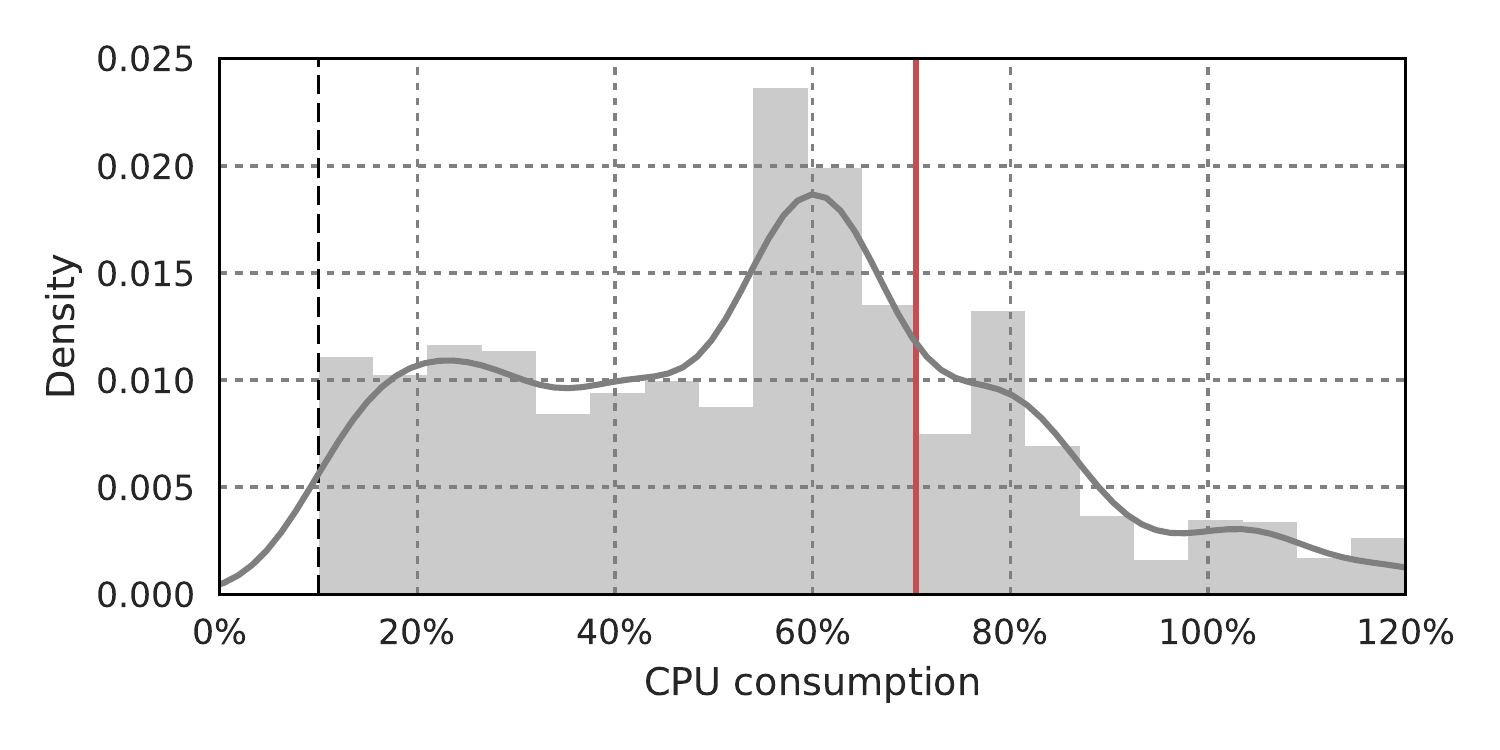}
  \vspace{-3mm}
  \caption{CPU consumption of \cryptojacking web pages.}
  \label{fig:greediness}
\end{figure}

The results of our analysis are shown in \fig{fig:greediness}. The most
popular setting is in the range \mbox{\num{50}--\perc{70}}.
Interestingly, about \perc{5}~of the websites attempt to even use up
more CPU cores than available on the system. This suggests that some
attackers are especially greedy in their attempts to max out the
available processing power by starting more mining threads than the CPU
can handle.

\subsection{Code Diversity}
\label{sec:clustering}

As the last step, we analyze the diversity of the JavaScript and
WebAssembly code in the identified miners. For JavaScript, we can
identify the script responsible for mining by inspecting the CPU
profiling. For the WebAssembly code, on the other hand, we have to
first concatenate all parsed functions into a single file using their
SHA1 hashes for ordering, as we separately obtain these functions from
the debugger. As a result of this preprocessing, we obtain one sample
of JavaScript and one merged sample of WebAssembly code for each of
the \num{1939} websites containing miners. While this representation
simplifies our experimental setup, it requires a fuzzy analysis, as
minor perturbations in the merged files obstruct the application of
exact matching.

\begin{figure}[htbp]
  \centering \subfigure[Similarity of JavaScript code]{
  \includegraphics[width=0.5\textwidth]{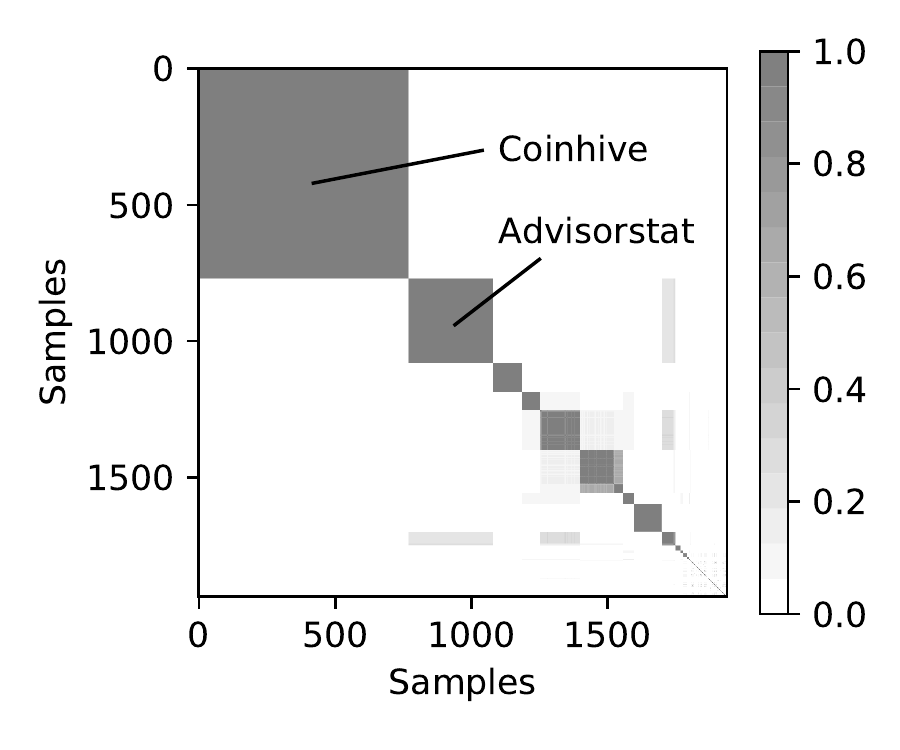}
}%
\subfigure[Similarity of WebAssembly code]{
  \includegraphics[width=0.5\textwidth]{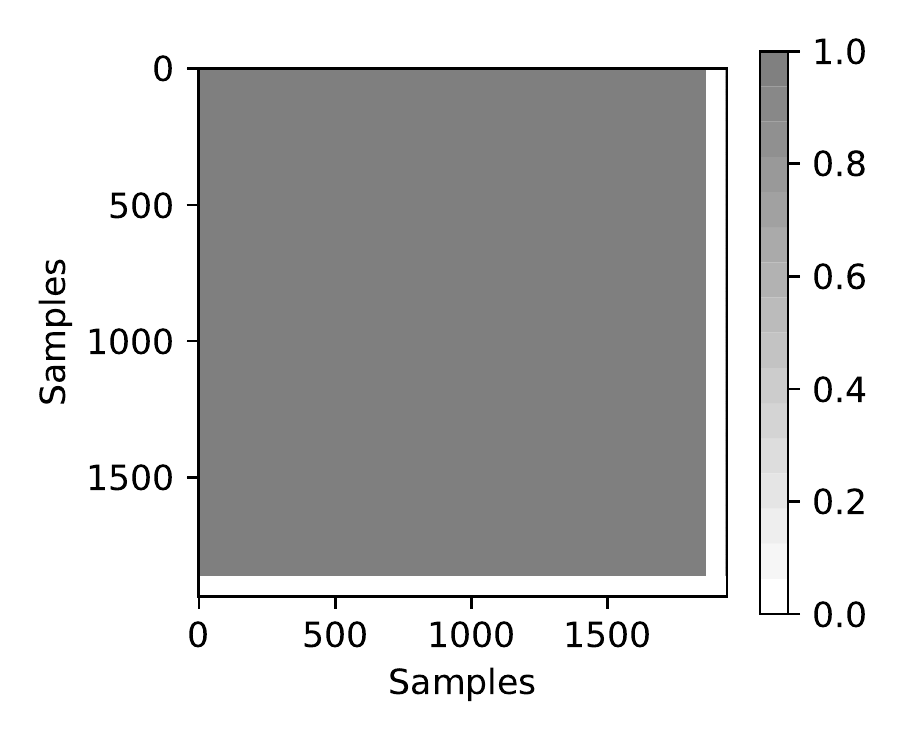}
  }\vspace{-3mm}
  \caption{Similarity and cluster analysis of JavaScript and
    WebAssembly code found in web-based miners.}
  \label{fig:simclust}
\end{figure}

We thus employ techniques from information retrieval that can cope
with noisy data. In particular, we conduct an $n$-gram analysis, where
the code samples are first partitioned into tokens using whitespaces
and then mapped to a vector space by counting the occurrences of
$n$-grams (sequences of $n$ tokens)~\citep[see][]{SalMcG86}. This
vectorial representation enables us to compute the \emph{cosine
  similarity} between all samples and generate the similarity matrices
shown in \fig{fig:simclust}. The columns and rows of the matrices are
arranged using hierarchical clustering, such that larger groups of
similar code samples become visible.

For JavaScript, we identify 23 clusters of similar code. These
clusters correspond to diverse implementations, including the
\coinhive library (largest cluster) as well as modified and obfuscated
code from different \cryptojacking campaigns. We discuss some of these
clusters in Section~\ref{sec:case-studies} as case studies. By
contrast, the WebAssembly code present in the \num{1939} websites
shows almost not diversity and is highly similar to the original
\coinhive implementation. We credit this finding to minor
modifications of the original code that allow for supporting
alternative currencies or operating over less expensive mining pools.

In summary, we conclude that the current landscape of cryptojacking is
dominated by variants of \coinhive. Although we spot different
JavaScript code wrapping the miners, the low-level code is in almost
all cases derived from a single implementation. Apparently, the
cryptographic primitives underlying the CryptoNote protocol and in
particular its proof-work-functions have only been once translated to
a web implementation and hence all active miners in our study rest on
the same foundation.

\section{Case Studies}
\label{sec:case-studies}

Finally, we briefly discuss the two largest families of miners that we
have identified in the previous section: 1)~\coinhive, the
implementation that has raised the most attention in public media, and
2)~Advisorstat, a lesser-known variant that employs a number of
techniques to obscure its operation.

\subsection{Case 1: CoinHive}

The \coinhive library first appeared in September~2017 on the German
image board ``pr0gram'', where it was tested as a payment and
incentive mechanism~\cite{website:krebs}. The implementation supports
different application scenarios, which range from replacing web
advertisments to rate-limiting traffic and monetizing services, such
as link shorteners and file hosters. Consequently, \coinhive shares
similarities with the framework developed by \citet{KarFraCap11} for
computation-based micropayments. It however differs from it by using
cryptocurrencies as target and thus directly
monetizing the computations.

Under the hood, \coinhive offers a flexible JavaScript API, with options
for throttling, not mining on mobile devices and opting-out. The latter
however is seldom presented to the user. For their services,
\coinhive currently charges a \perc{30} fee. In our dataset, we found
\num{836} sites that directly include the mining code from
\code{coinhive.com/lib/coinhive.min.js}. However, we observed \num{940} sites
that communicate with \code{ws*.coinhive.com/proxy}, the WebSocket
backend of \coinhive, without previously requesting the mining code from
\coinhive's servers. This shows that some sites prefer to host the script on
their own servers or retrieve it from other sources, while still using
\coinhive's infrastructure for the actual mining. A likely reason for this
behavior is an attempt to hide the presence of the mining activity, for
instance, to circumvent blacklist-based countermeasures that check for the
original \coinhive-script URL.

\subsection{Case 2: Advisorstat}

The second biggest cluster is active on \num{315} different sites,
all of which use a mining script hosted at \code{advisorstat.space/js/advisor.js}. 
A direct visit to \code{advisorstat.space} results in a \code{403 
Forbidden} message from Nginx. Apparently, the mining script and its
Wasm counterpart are the only content hosted on the domain. This
raises the question why many, seemingly unrelated sites make use of the
same unknown and unadvertised mining service. Further investigation of
the affected sites reveals the common link: All sites in this cluster
are hosted by the same provider for free website creation called
\textit{uCoz}. The script is delivered via a banner at the top of each
site, which advertises uCoz's services.

During further examination of the affected sites, we have quickly noticed that the
miner only is active on the first visit of each affected site. The reason for
this is a cookie called \code{uclickadushowed}, which is created with an
expiration date of \num{12}~hours upon visiting any site with the aforementioned
banner. Interestingly, the presence of this cookie is checked on the
server-side and the code responsible for the miner is not included in the
response, if the cookie exists. Following the trail of the banner,
we observe that first obfuscated JavaScript code is loaded from
\code{moradu.com}, which subsequently loads further code from \code{netrevgo.com}.
Both these domains are registered through \code{privacyprotect.org} and
hosted on Amazon's AWS. Finally, the mining code is retrieved from
\code{advisorstat.space}, a domain registered in Panama through
\code{whoisguard.com}, another privacy protection service. The mining script
itself contacts the advisorstat server for reporting the hashes using obfuscated packets sent
over a WebSocket connection.

\section{Discussion and Limitations}
\label{sec:limitations}

Our study provides the first comprehensive view on cryptojacking in
the wild. Nonetheless, several of our results are estimates and
approximations, as exact measurements are hardly possible in a dynamic
system such as the Internet. In the following, we discuss the
implications and limitations of our findings in more detail and reason
about their practical relevance.

\subsection{Revenue}

Determining the overall revenue of \cryptojacking is inherently
difficult, due to the lack of exact measurements for the number of website
visits or their duration. The popularity of a website, its content
and, furthermore, its potential to engage visitors to stay are subject
to constant change. In this study, we hence operate on estimates of
the profit only and the ascertained values are only as accurate as the
visiting statistics we use.

We show that web-based mining can generate significant revenue for
large websites. The video-streaming platform \pornhub, for instance,
may earn as much as \SI{50000}{USD} a day using the \coinhive
service. This can be even increased if a mining service with a less
severe payout ratio is used. However, considering traditional online
advertisements with a typical payout of \SI{1}{USD}~per thousand
impressions (CPM), \pornhub, for instance, would already make
\SI{81000}{USD} per day. On this level, \cryptojacking thus does not
seem to be a worthwhile replacement. For websites with fewer visits
from countries for which advertisement networks do not pay as well the
ratio may shift.  Our study, however, shows that the average revenue
only is moderate.

\subsection{Countermeasures}

As demonstrated, \cryptojacking is a prevalent malice in today's web
landscape. Thus, it is of interest to investigate suitable
countermeasures to protect web users from the parasitic usage of their
resources. The evaluation in Section~\ref{sec:blacklists} has shown
that existing blacklist-based approaches are ineffective, as they are
trivial to evade and the actual lists outdate fast.

Instead of static blacklists, in this paper, we leverage a set of
heuristic indicators for candidate selection and a dedicated
performance measurement step for precise miner identification
(see~Section~\ref{sec:identification}). As shown, this approach is
well suited to reliably detect the current generation of mining
scripts. However, this has to be partially attributed to the fact,
that today's mining operators apparently do not anticipate our
detection approach.

Unfortunately, none of our utilized heuristic indicators constitute a
necessary technical precondition to implement mining JavaScript code:
The communication via WebSockets can be replaced with other JavaScript
networking capabilities, such as \texttt{XMLHttpRequest}. The
parallelization of script execution through WebWorkers can be imitated
through spawning multiple \texttt{iframes}---even though this
technique leads to the loss of active mining in non-focused browser
tabs. And while there is no real alternative to WebAssembly, less
performant miner scripts can be implemented using subsets of JavaScript, such
as \texttt{asm.js}.

As a result, the only reliable indicator in the presence of an
adversary that actively tries to avoid detection is the measurement of
prolonged and excessive CPU usage. Simple CPU thresholds carry the
danger of false positives in case of websites with occasional CPU
heavy tasks or potential false negatives in the form of mining scripts,
which deliberately try to remain below a certain CPU usage limit.
Instead, we advocate the exploration of \textit{CPU allotments}, in
which each browser tab receives a certain amount of \textit{CPU
  minutes}. As soon as a tab runs out of its quota, the browser could
take actions, such as throttling the tab's scripts or warning the
user. We leave an implementation and evaluation of such a mining-aware
browser to future work.

\subsection{Threats to Validity}

Finally, we discuss threats to the validity of our empirical study and
how they have been addressed in our implementation and experimental
setup. Moreover, we pinpoint directions for extensions that can limit
certain effects in future studies.

\paragraph{False positives}

The core of this paper consists of a detection approach that aims to
find \cryptojacking scripts at run-time. This task comes with an
inherent precision problem, as we try to determine the semantics of
executed code from dynamic execution artifacts based on a set of
heuristics and characteristics. Thus, even though the used set of
indicators and heuristics are carefully chosen, based on thorough
manual analysis of validated \cryptojacking code, there are no formal
guarantees that the approach really identifies mining scripts or does
not accidentally misclassify some examined sites, for instance, due to
computation-heavy, legitimate JavaScript code.

Creating a good testbed for false positives is unfortunately impossible: Simply
hosting a set of web applications from trustworthy sources ourselfes would not
achieve much, as widely available software is likely to be well tested and free
of obvious performance issues. Rather, we would expect a high CPU load on a
custom, hand-crafted website built by an inexperienced developer, who
unintentionally uses very ineffectient code to interact with the DOM or
introduces some kind of endless loop. As there does not exist a set of
real-world, buggy-but-benign websites, we can not test our system for false
positives on a ground truth.

To examine the potential problem of misclassification, we instead use a
similarity analysis on the resulting data set of JavaScript and Wasm code (see
Section~\ref{sec:clustering}). While in the collected set of JavaScript a
certain degree of variance exists, the vast majority of Wasm code exhibits an
astonishingly high degree of similarity, with less than 4\% of outliers.
To further investigate the JavaScript code, we took a random sample from each
of the 23 clusters to represent that cluster. By combining manual static and
dynamic analysis and searching for mining-specific strings and functions, we
were able to confirm that each sample is indeed a cryptominer and thus the
whole cluster is likely to contain only scripts used for \cryptojacking.

\paragraph{False negatives}

The study only provides a lower bound on the overall \cryptojacking
landscape, as we are aware of a set of scenarios, which are currently
not covered by our methodology: First, we only visit the homepages
of the Alexa Top 1 million without deeper crawling of the sites. Thus,
if a website uses miners exclusively in some of its subpages, our
crawler would not have encountered them. Second, the detection process
might miss mining sites that deliberately delay the inclusion of the
mining scripts in the web document for a time that exceeds our
analysis time window. Similarly, sites that use a non-deterministic
condition to start the mining process or require user interaction for
the mining to start will not be found. Finally, custom, non-active
mining scripts for which no static indicators can be generated during
phase~3 are missed by our approach.

To address this issues, we did a thorough cross-check with the current
state-of-the-art in miner detection, namely the most used browser extensions.
Our approach achieves high coverage of the manually curated blacklists of the
extensions (see Section~\ref{sec:blacklists}), which conclusively shows our
technique's ability to identify mining scripts. Furthermore, increasing the
reach of the crawling process and extended examination times would in general
address the majority of the potential problems. Also, periodic repetition of
the experiment will lead to the eventual detection of unreliable or currently
malfunctioning miners. We leave these measures to future work.

Regarding the use of evasions to prevent detection, it should be noted
that techniques like delaying the start of the miner also come with a
significant drawback: Unlike traditional malware infections, where the malware
likely can achieve persistance on the system, web-based mining stops as soon as
the user closes the browser tab. Thus, an attacker only has limited time to run
the mining code and any delays will negativly affect his profits, making it less
likely to encounter such techniques.

\paragraph{Data analysis}

The results of the website distribution measurements
(Section~\ref{sec:distribution}) and revenue estimations
(Secion~\ref{sec:revenue}) directly rely on external data sources,
such as GeoIP, SimilarWeb, or Symantec’s WebPulse Site Review. Thus,
the quality of the provided analysis depends on the quality of the
external data. Especially, the revenue calculations rely on
\textit{estimated} figures that are compiled using proprietary
methodologies. Thus, the impact of potential problems in the
underlying data should be considered when interpreting the presented
results.

\section{Related Work}
\label{sec:related-work}

Web-based \cryptojacking is a novel attack strategy that has received
little attention in the research community so far. While news media
and web blogs have started to cover different incidents involving
\cryptojacking in the last months~\citep[\eg][]{website:krebs,
  website:adguard, website:degroot, website:goodin17}, little systematic
investigation of the prevalence and efficacy of this threat has been
conducted to date. The study by \citet{EskLeoMurCla18} was the first to provide
a peek at the problem. However, the study is limited to vanilla \coinhive
miners, and the underlying methodology is unsuited to detect alternative or
obfuscated mining scripts. In contrast, our work aims to provide a
comprehensive survey on the landscape of \cryptojacking in the web using a
technology-agnostic detection approach. 

In independent, concurrent work, \citet{MineSweeper} search the web for
instances of drive-by mining. They first identify miners with a list of static
keywords in the JavaScript code and additionally utilize dynamically collected
data in the form of WebSocket communication and the number of created
WebWorkers. In March 2018 they crawled the Alexa Top 1 Million front-pages plus
three sub-pages for each and identified \num{1735} websites with a miner. Building on
these results, they propose a detection based on the identification of
cryptographic primitives inside the Wasm code. This alternative detection
dubbed MineSweeper found \num{744} miners in the Top 1 Million (without
sub-pages) in their crawl in April. Similarly, we intinally also use a
combination of static and dynamic indicators to identify websites that
\emph{might} include a miner, but in contrast then use V8's profiler to measure
the CPU usage on a per-function level over an extended timeframe to
\emph{verify} the persence of miner. With this approach, we find \num{2506}
websites conducing \cryptojacking in the Alexa Top 1 million without visiting
any sub-pages.

Also related is the work of \citet{KarFraCap11} that investigate benign
applications of computation-based micropayments in the web. While their work
lays out the positive potential of microcomputations, such as mining, our study
focuses on the current abuse of this concept and its occurrence in the web.
Unauthorized mining of cryptocurrencies, however, is not limited to web
scenarios. For example, \citet{Huang2014} present a study on malware families
and botnets that use Bitcoin mining on compromised computers. Similarly,
\citet{Ali2015} investigate botnets that mine alternative currencies, such as
Dogecoin, due to the rising difficulty of profitably generating Bitcoins.  To
detect illegitimate mining activities, either through compromised machines or
malicious users, \citet{Tahir2017} propose \emph{MineGuard}, a hypervisor-based
tool that identifies mining operations through CPU and GPU monitoring. Our
study extends this body of work by providing an in-depth view of mining
activity in the web.

From a more general point of view, cryptocurrency mining is a form of
\emph{parasitic computing}, a type of attack first proposed by
\citet{parasitic:nature01}. As an example of this attack, the authors
present a sophisticated scheme that tricks network nodes into solving
computational problems by engaging them in standard
communication. Moreover,
\citet{ParraRodriguez:2018:CCS:3176258.3176951} present an alternative
method for abusing web technology that enables building a rogue
storage network. Unlike \cryptojacking, these attack scenarios are
mainly of theoretical nature, and the authors do not provide evidence
of any occurrence in the wild.

On a technical level, our methodology is related to approaches using
high-interaction honey browsers \citep[\eg][]{Provos2008, Wang2006,
  Moshchuk2007, KolLivZorSei12}, which are mainly utilized to detect
attacks on the browser's host system via the exploitation of memory
corruption flaws, a threat also known as
\textit{drive-by-downloads}. While our approach shares the same
exploration mechanism---using a browser-like system to actively visit
potentially malicious sites---our detection approach diverges, as the
symptoms of browser-based mining stem from the exclusive usage of
legitimate functionality, in contrast to drive-by-download attacks
that cause low-level control-flow changes in the attacked browser or
host system.

\section{Conclusion}
\label{sec:conclusion}

This study provides the first comprehensive view on the threat of
web-based cryptojacking. Although browser-based mining is a rather novel
development, our empirical investigation reveals an increasing number of
websites that employ this technology and exploit computational resources
of their visitors. We show that approximately
\num{1}~out~of~\num{500}~websites in the Alexa 1 million ranking
contains a miner that immediately starts mining when visiting the
website. This implies that falling victim to a
\cryptojacker is not a rare event, and a considerable amount of energy
is drained as part of this illegal activity every day.

Despite several mining websites with thousands of visitors, our
estimate of the generated revenue shows that web-based \cryptojacking
is not as profitable as it seems and many miners attain only moderate
payouts. Still, cryptocurrencies enjoy great popularity and provide a
lucrative playground for financial speculation. It is thus unlikely
that mining activity will disappear on its own unless cryptocurrencies
significantly loose in value or novel regulations limit their
trading. As a consequence, there clearly is a need for effective
detection and defense mechanisms.

Unfortunately, we show in our study that current detection mechanisms
are insufficient to fend off this threat, as they rely on simple
blacklists that fail to cope with the complexity of JavaScript and
WebAssembly code. This complexity can only be tackled if defense
mechanisms are tightly integrated into the browser, such that the
resources available to a website can be monitored and regulated
dynamically---irrespective of the execution environment and employed
web standards. Ultimately, such protection might help to generally
mitigate the threat of parasitic computing inherent to current web
technology.

\section*{Acknowledgments}
The authors would like to thank Martina Lindorfer and Herbert Bos for providing a draft of their related paper. Furthermore, the authors gratefully acknowledge funding from the German Federal Ministry of Education and Research (BMBF) under the project VAMOS (FKZ 16KIS0534) and FIDI (FKZ 16KIS0786K) and funding from the state of Lower Saxony under the project Mobilise.

\bibliographystyle{abbrvnat}
\balance
\bibliography{goldrush}

\begin{backpage}
\showtubslogo
\begin{titlerow}[bgcolor=\clrDark,fgcolor=tubsWhite]{3}
  \large
  Technische Universität Braunschweig\\
  \appsec\\
  Mühlenpfordtstraße 23\\
  38106 Braunschweig\\
  Germany
\end{titlerow}

\begin{titlerow}[bgcolor=\clrLight]{4}
  \large
  ~
\end{titlerow}
\end{backpage}

\end{document}